\documentclass[aps,prb,twocolumn]{revtex4}
\usepackage{graphicx}
\pdfoutput=1

\usepackage{amssymb}
\usepackage{graphicx}
\usepackage{amsmath}
\usepackage{graphicx}
\usepackage{dcolumn}
\usepackage{bm}
\usepackage{color}

\begin{document}

\title{Band structure and electron-phonon coupling in H$_3$S:
a tight-binding model}

\author{L. Ortenzi, E. Cappelluti, and L. Pietronero}

\affiliation{Istituto dei Sistemi Complessi - CNR, 00185 Roma, Italy}

\affiliation{Dipartimento di Fisica, Universit\`a ``La Sapienza'',
P.le Aldo Moro 2, 00185 Roma, Italy}

\date{\today}

\begin{abstract}
We present a robust tight-binding description,
based on the Slater-Koster formalism, of the band structure of H$_3$S
in the {\em Im}$\bar{3}${\em m} structure, stable in the range of
pressure $P = 180-220$ GPa.
We show that the interatomic hopping between the 3$s$ and 3$p$ orbitals
(and partially between the 3$p$ orbitals themselves)
of sulphur
is fundamental to capture the relevant physics associated with the Van Hove
singularities close to the Fermi level.
Comparing the model so defined with density functional theory calculations
we obtain a very good agreement not only of the overall band-structure,
but also of the low-energy states and of the Fermi surface properties.
The description in terms of Slater-Koster parameters permits us also to evaluate
at a microscopic level a hopping-resolved linear electron-lattice coupling
which can be employed for further tight-binding analyses also at a local scale.
\end{abstract}

\maketitle

\section{Introduction}

The recent report of high-temperature superconductivity with $T_c > 200$ K 
in compressed sulphur hydride opens new exciting perspectives in condensed matter
and in the wider field of physics.~\cite{Eremets_nature}
The practical applications of this superconducting state is however
hampered by the need of extremely high pressures $P$.
Within this context a microscopic understanding of the underlying
electronic states responsible for the pairing, and of many-body 
interaction mechanisms is thus fundamental to master the origin 
and properties of such extremely high critical temperature, 
with the perspective to predict new materials or suitable conditions 
where high-$T_c$ superconductivity can be achieved at
room pressure.\cite{ph3}

Most of the theoretical work in compressed sulphur hydrides have been
carried out so far by means of first-principle calculations, providing
useful insights.\cite{Igor,Mauri,Duan1,Duan2,Gross,Pickett,Pickett2,komelj,Akashi2015,boeri,ge,li,bianconi1,bianconi2,errea2} 
For instance, the most stable compound in the range of pressure relevant 
for superconductivity ($P > 180$ GPa) has been predicted to be H$_3$S
with {\em Im}$\bar{3}${\em m} structure.~\cite{Duan1,Igor,Mauri}
Such prediction has been also confirmed by recent X-ray experiments.\cite{li,einaga,goncharov}
The band structure and the electron-phonon coupling have been also
computed by means of ab-initio techniques.~\cite{Igor,Mauri,Duan1,Duan2,Gross,Pickett,Pickett2,Akashi2015,boeri,ge,li,bianconi1,bianconi2} 
More detailed calculations show that the high-frequency vibrational mode
can be concomitant with a large coupling $\lambda \sim 2$,
whose origin is however not strictly related to the vibrational spectrum,
but it can also profit of a locally high density of states $N(0)$ at the
Fermi level.~\cite{Gross,Pickett,Pickett2}
Such high density of states, on the other hand, has been, associated
with the presence of two Van Hove singularities very close to the
Fermi surface.\cite{Pickett2,bianconi1,bianconi2}
Along with density functional theory (DFT) calculations,
a couple of tight-binding models have been also discussed
in the field,\cite{Igor,Pickett2} but without succeeding in
reproducing the Fermi surfaces and the Van Hove singularities,
and hence the main electronic characteristics of these compounds.

Although the overall scenario provided by first-principle
calculations suggests the picture that H$_3$S can be one
of the best optimized conventional superconductor,\cite{carbotte,Igor-Nature,dura}
the actual experimental phenomenology of this compound is quite non
trivial.
Raman spectra for instance are extremely broad and not conclusive,\cite{Eremets_nature} as likely results of the large atomic zero point motion. 
The relevance of such quantum fluctuations has been also pointed out
in Refs. \onlinecite{bianconi1,bianconi2,errea2,banacky,sano}.
Furthermore,
the superconducting critical temperature has been shown to depend crucially
on the annealing processes and on the pressure/cooling procedures.\cite{Eremets_nature}
These observations suggest that lattice fluctuations and local effects
due to interfaces, defects and impurities, and to the coexistence
of different metastable phases,\cite{akashi2}
 can possibly affect and control the
superconducting properties. Addressing these issues in first-principle
approaches is however a hard task, because of the need of a large
supercell.
A reliable tight-binding model is thus highly desired, with the aim
to address all these issues with a affordable computational cost,
and to provide an analytical description of the elementary excitations
and of their coupling with the vibrational degrees of freedom,
which can be easily extended at the local level.


 In this work , using ab-initio calculations as a reference, we present
a suitable Slater-Koster tight-binding model, providing an accurate 
analytical description of the electronic band structure.
We show that the Van Hove singularities can be properly reproduced
by taking into account the role of the hopping between nearest
neighbor $s$ and $p$ sulphur orbitals.
The description of the tight-binding model in terms of a
Slater-Koster approach,~\cite{S-K} in addition, allows
for an analytical description of the electron-phonon coupling
at the local as well as a uniform level.
The strength of the electron-phonon coupling is thus related to the
dependence of the Slater-Koster hopping parameters on the interatomic distance.
The hopping-resolved electron-phonon coupling is hence numerically
computed by using first-principle calculations.

The tight-binding model here presented can provide a paradigmatic base
for the several future developments, as for instance the investigation
at the local scale of impurities, vacancies, hopping disorder and
grain boundaries.
Lattice distortions can be as well included at the classical level.
The present tight-binding model represents also a suitable basis
for the inclusion of many-body (electron-electron and electron-phonon)
effects within the Quantum Field Theory framework.\cite{sano}

\section{The model}

\begin{figure}
\begin{center}
\includegraphics[angle=0, width=6cm]{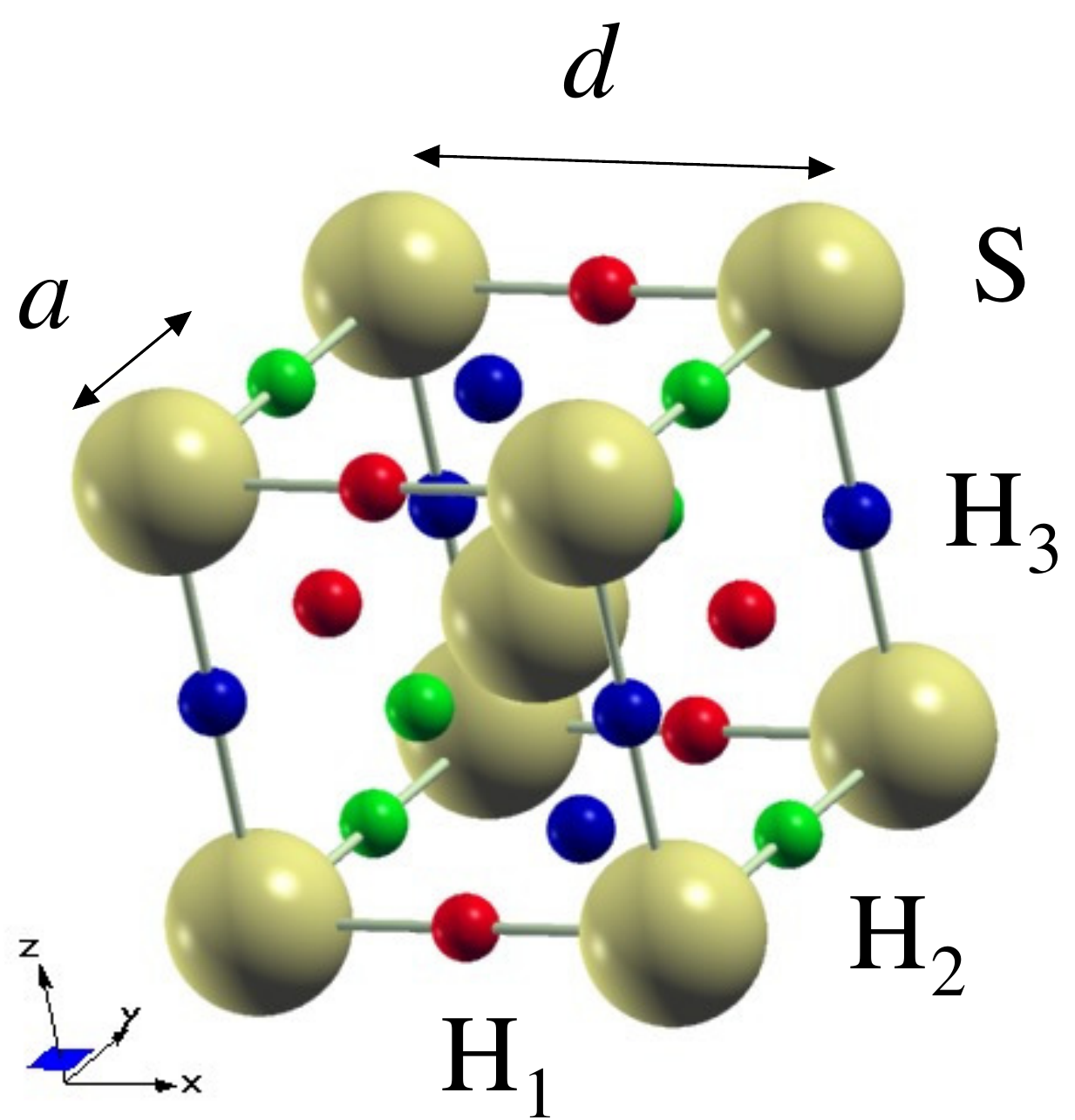}
\caption{(Color online) Lattice structure of H$_3$S. $a$ denotes the
shortest H-H distance (or, equivalently, the shortest H-S distance,
$d=2a$ is the size of the cubic structure of the S bcc lattice.}
\label{f-crystal}
\end{center}
\end{figure}

We investigate H$_3$S in the range $P=180-220$ GPa, where
the material has predicted to a structure belonging
to the {\em Im}$\bar{3}${\em m} space group,\cite{Duan1,Duan2}
with the sulphur atoms lying on a bcc lattice, and three hydrogen
atoms per unit cell on the octahedral sites around each sulphur.
The crystal structure is shown in Fig. \ref{f-crystal}, where
we denote $a$ the nearest-neighbor H-H (or equivalently H-S) distance.
For our convenience, we also label H$_1$, H$_2$, H$_3$ the three hydrogen atoms
per unit cell according Fig. \ref{f-crystal}.
The volume change as a function of the pressure has been studied
in Ref. \onlinecite{Duan1,Duan2},
varying from $a=1.5075$ \AA\, at $P=180$ GPa, to
$a=1.4795$ \AA\, at $P=220$ GPa.
The Brillouin zone is thus characterized by the high-symmetry points
$\Gamma$=(0,0,0), H=($\pi/a$,0,0), N=($\pi/2a$,0,$\pi/2a$),
P=($\pi/2a$,$\pi/2a$,$\pi/2a$).
Also relevant will appear the point F=($3\pi/4a$,$\pi/4a$,$\pi/4a$)
with lower symmetry, which lies midway between H and P.

\begin{figure}[t]
\begin{center}
\includegraphics[angle=0, width=8cm,clip=]{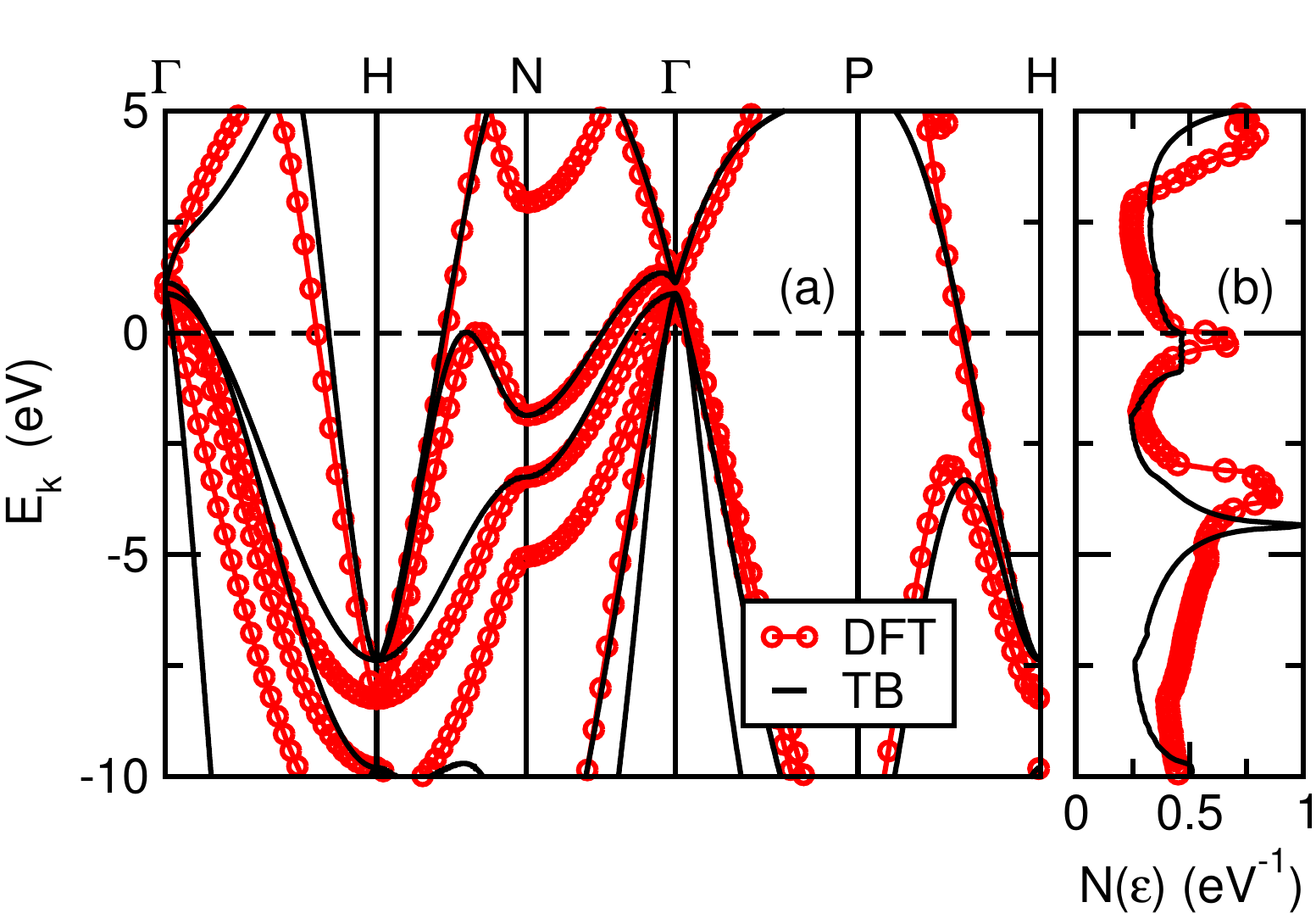}
\includegraphics[angle=0, width=7cm,clip=]{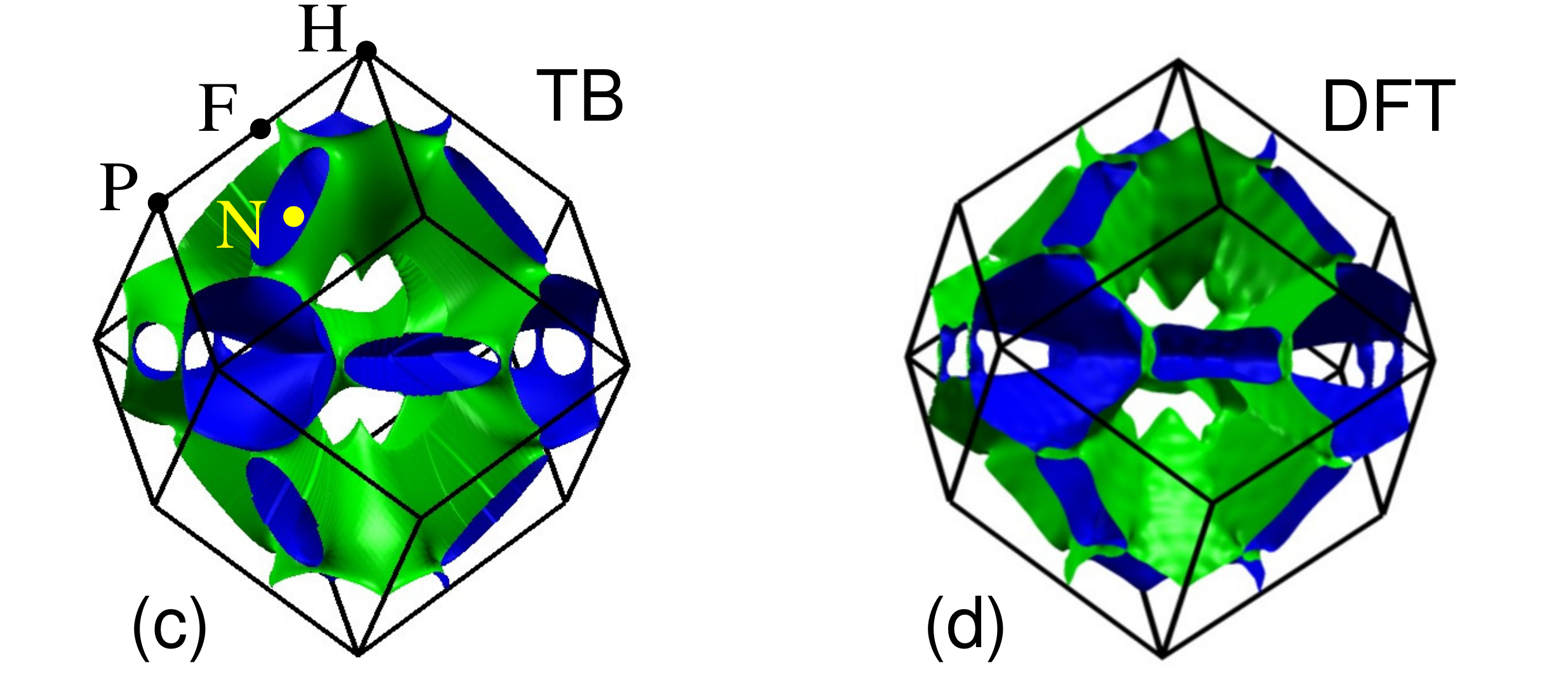}
\caption{(Color online) Panel (a): first-principle band structure of H$_3$S (red
dots), compared with the band structure resulting from the
tight-binding model (black solid lines). 
Panel (b): corresponding density of states $N(\epsilon)$.
The black dashed lines mark the position of the Fermi level E$_{\text{F}}$.
The comparison of the Fermi surfaces of the most relevant band
obtained from first-principle calculations and from our tight-binding model
is shown in panels (c)-(d).
}
\label{f-bands}
\end{center}
\end{figure}

Density-functional-theory calculations were performed
using the generalized gradient approximation (GGA),
with the linear augmented plane wave (LAPW) method as implemented in the
\textsc{wien2k} code.~\cite{WIEN2K,DFT:PBE}
Up  to  288 $\mathbf{k}$ points were used in the self-consistent
calculations with an LAPW basis defined by the cutoff $R_SK_{max}=6$.
A larger number of 2456 $\mathbf{k}$ points were used for calculating
the density of states (DOS).
The resulting electronic band structure along the relevant axes of
high-symmetry $\Gamma$-H-N-$\Gamma$-P-H, for the representative case
$P=200$ GPA, is shown in Fig. \ref{f-bands}a,
whereas the corresponding electronic density of states $N(\epsilon)$
is shown in the panel (b).\cite{notefermilevel}
Both calculations are in agreement with the previously reported
first-principle calculations.~\cite{Igor,Mauri,Duan1,Duan2,Gross,Pickett,Pickett2,Akashi2015,bianconi1,bianconi2}
Most noticeable is the peak in the density of states,
which arises from two (upper and lower) Van Hove singularities (VHs)
close to the Fermi level.~\cite{Pickett,Pickett2}
The upper VHs is visible in the band structure [Fig. \ref{f-bands}a]
as a local maximum along the cut H-N, whereas the lower VHs lies
out of the high-symmetry cuts and it is not immediately visible
in Fig. \ref{f-bands}a\cite{Pickett2}.

As mentioned in many works, the orbital character of the band structure close to
the Fermi level is mainly related to the $1s$ orbitals of hydrogen,
and to the $3s$, $3p$ orbitals of sulphur.
Following Refs. \onlinecite{Igor,Pickett2}, we restrict thus our analysis to the
Hilbert space defined by the 7-fold vector
\begin{eqnarray}
\phi_i^\dagger
&=&
(
h_{i,1}^\dagger,
h_{i,2}^\dagger,
h_{i,3}^\dagger,
s_{i}^\dagger,
p_{i,x}^\dagger,
p_{i,y}^\dagger,
p_{i,z}^\dagger
),
\label{basis}
\end{eqnarray}
where $h_{i,\alpha}^{\dagger}$ creates an electron on the $\alpha$-labeled H atom
in the $i$-unit cell, $s_{i}^{\dagger}$ creates an electron in the $3s$ orbital
of the S atom in the $i$-unit cell, and $p_{i,\alpha}^\dagger$ creates
an electron in the $3p_\alpha$ orbital of the S atom in the $i$-unit cell.
In order to provide a tight-binding model as a suitable basis for
the inclusion of many-body effects by means of diagrammatic techniques,
we assume that the basis orbitals are orthonormal, so that the overlap
matrix is the unit matrix.

A tight-binding Hamiltonian is defined by the specification of all
the on-site orbital energies and of all the relevant hopping terms.
A basic model was introduced in Ref. \onlinecite{Igor},
were only hoppings between nearest neighbors H-H and H-S
(namely, with inter-atomic distance $a$)
were considered.
Such model was further improved in Ref. \onlinecite{Pickett2},
including further (selected) interatomic hoppings,
but still being unable to reproduce the fundamental features
related to the Van Hove singularities.

As we show see below, a crucial ingredient missing in the previous
tight-binding models is the direct hopping between $3s$ and $3p$ 
orbitals on nearest  neighbor sulphur atoms.
In our analysis,  we include thus, using a  Slater-Koster framework, the nearest neighbor
hybridization between {\em all} the orbitals of the Hilbert space.
More explicitly we consider:
the hopping between nearest $1s$ orbitals of the hydrogen,
at distance $a$, ruled by the Slater-Koster parameter $H_{ss\sigma}$;
the hopping between nearest $3s$ orbitals of the sulphur,
at distance $\sqrt{3}a$, ruled by the parameter $S_{ss\sigma}$;
the hopping between nearest $3p$ orbitals of the sulphur,
at distance $\sqrt{3}a$, ruled by the Slater-Koster parameters
$S_{pp\sigma}$, $S_{pp\pi}$;
the hopping between the $1s$ hydrogen orbital
and the $3s$ sulphur orbital, on nearest neighbor H-S atoms
at distance $a$, governed by the parameter $U_{ss\sigma}$;
the hopping between the $1s$ hydrogen orbital
and the $3p$ sulphur orbital, on nearest neighbor H-S atoms
at distance $a$, governed by the parameter $V_{sp\sigma}$;
and the hopping between the sulphur orbitals $3s$ and $3p$
on nearest neighbor S atoms at distance $\sqrt{3}a$
tuned by the parameter $W_{sp\sigma}$.
Additional parameters of the tight-binding model are the on-site
orbital energies $\epsilon_{\rm H}$, $\epsilon_{\rm S_s}$,
$\epsilon_{\rm S_p}$, corresponding to the H $1s$,
S $3s$, S $3p$, respectively.

The Hamiltonian matrix assumes thus the relatively simple form:
\begin{eqnarray}
\hat{H}({\bf k})
&=&
\left(
\begin{array}{ccc}
\hat{H}_\sigma({\bf k})
& \hat{U}_\sigma({\bf k}) 
& \hat{V}_\sigma({\bf k})  \\
\hat{U}_\sigma^\dagger ({\bf k})
& \hat{S}_\sigma({\bf k})
& \hat{W}_\sigma ({\bf k}) \\
\hat{V}_\sigma^\dagger ({\bf k})
& \hat{W}_\sigma^\dagger ({\bf k})
& \hat{S}_\pi({\bf k})
\end{array}
\right),
\label{hmatr}
\end{eqnarray}
where
\begin{eqnarray}
\hat{H}_\sigma({\bf k})
&=&
\left(
\begin{array}{ccc}
\epsilon_{\rm H} & 2 H_{ss\sigma} C_z & 2 H_{ss\sigma} C_y \\
2 H_{ss\sigma} C_z & \epsilon_{\rm H} & 2 H_{ss\sigma} C_x \\
2 H_{ss\sigma} C_y & 2 H_{ss\sigma} C_x & \epsilon_{\rm H}
\end{array}
\right),
\label{hsmatr}
\end{eqnarray}
\begin{eqnarray}
\hat{S}_\sigma({\bf k})
&=&
\epsilon_{\rm S_s}
+
8S_{ss\sigma}C_{x,y,z},
\end{eqnarray}
\begin{eqnarray}
\hat{S}_\pi({\bf k})
&=&
\epsilon_{\rm S_p}
\hat{I}
+
\nonumber\\
&&
\frac{8}{3}\left(
\begin{array}{ccc}
S_{pp}^+C_{x,y,z}
& S_{pp}^-S_{x,y}C_z
& S_{pp}^-S_{x,z}C_y
\\
S_{pp}^-S_{x,y}C_z
& S_{pp}^+C_{x,y,z}
& S_{pp}^-S_{y,z}C_x
\\
S_{pp}^-S_{x,z}C_y
& S_{pp}^-S_{y,z}C_x
& S_{pp}^+C_{x,y,z}
\end{array}
\right),
\end{eqnarray}
\begin{eqnarray}
\hat{U}_\sigma({\bf k})
&=&
2U_{ss\sigma}
\left(
\begin{array}{c}
C_x \\
C_y \\
C_z
\end{array}
\right),
\end{eqnarray}
\begin{eqnarray}
\hat{W}_\sigma({\bf k})
&=&
i\frac{8W_{sp\sigma}}{\sqrt{3}}
\left(
\begin{array}{ccc}
S_xC_{y,z} & S_y C_{x,z} & S_z C_{x,y}
\end{array}
\right),
\end{eqnarray}
\begin{eqnarray}
\hat{V}_\sigma({\bf k})
&=&
2iV_{sp\sigma}
\left(
\begin{array}{ccc}
S_x & 0 & 0 \\
0 & S_y & 0 \\
0 & 0 & S_z
\end{array}
\right),
\label{vmatr}
\end{eqnarray}
and where
$S_{pp}^+=S_{pp\sigma}+2S_{pp\pi}$,
$S_{pp}^-=S_{pp\sigma}-S_{pp\pi}$,
$C_i=\cos(k_ia)$, $S_i=\sin(k_ia)$,
$C_{i,j}=\cos(k_ia)\cos(k_ja)$,
$S_{i,j}=\sin(k_ia)\sin(k_ja)$,
$C_{i,j,l}=\cos(k_ia)\cos(k_ja)\cos(k_la)$.

Eqs. (\ref{hmatr})-(\ref{vmatr}) define
our tight-binding Hamiltonian in terms
of ten energy parameters:
$\epsilon_{\rm H}$, $\epsilon_{\rm S_s}$, $\epsilon_{\rm S_p}$,
$H_{ss\sigma}$, $S_{ss\sigma}$, $S_{pp\sigma}$, $S_{pp\pi}$,
$U_{ss\sigma}$, $V_{sp\sigma}$, $W_{sp\sigma}$.

The matrix structure of such Hamiltonian is simple enough to allow
for an analytical solution in all the high-symmetry points
$\Gamma$, H, N, P,
of the Brillouin zone.
A detailed analysis of the eigenvalues
of the Hamiltonian in these high-symmetry points
is provided in Appendix \ref{a:eigen}.
A careful inspection of the orbital character permits also
to identify qualitatively each eigenstate of the TB Hamiltonian
with a corresponding level in the DFT band structure.
It is clear that, given the limited number (ten) of adjustable
tight-binding parameters, it is not possible to match the energy
of each DFT level with a corresponding tight-binding eigenvalues
on all the high-symmetry points.
Moreover, a careful inspection reveals that
the parameter $W_{sp\sigma}$,
as well as the combination  $S_{pp}^-$,
{\em never} appear
at the high-symmetry points $\Gamma$, H, N, P.\cite{note_TB}
However, as we show below, the parameters $W_{sp\sigma}$ and $S_{pp}^-$
play a pivot role in the realistic band structures,
being responsible for the non-monotonic dispersion of the conduction band
along the cut H-N, whose maximum is associated with the most relevant Van Hove singularity close to the Fermi level.

In this regards, we can explicitly compare the analytical properties of
our tight-binding model with the ones presented in
Refs. \onlinecite{Igor,Pickett2}.
In particular, we notice that the model of Ref. \onlinecite{Igor} can
be obtained
by considering only the hopping terms $H_{ss\sigma}$, $U_{ss\sigma}$ and
$V_{sp\sigma}$.
On the other hand, the minimal model of Ref. \onlinecite{Pickett2}
corresponds to consider the nearest neighbor hoppings
$U_{ss\sigma}$, $H_{ss\sigma}$, the linear combination
$S_{pp}^+$, plus few farther neighbor hoppings.
In both the cases, the hoppings
$W_{sp\sigma}$ and $S_{pp}^-$ were not included.

\section{Band structure and tight-binding parameters}

Eqs. (\ref{hmatr})-(\ref{vmatr}) provide us a suitable tool
to describe in an accurate way the band structure of H$_3$S.
To this aim, guided by DFT calculations, the ten tight-binding parameters must be specified.
A widely used procedure, employed in Refs. \onlinecite{Igor,Pickett2},
is to evaluate the hybridization constants between atomic orbitals
on different locations by means of a Wannier representation.
This method, however, as mentioned in Ref. \onlinecite{Pickett2},
is not straightforwardly cast in terms of a Slater-Koster description.
In order to show the feasibility of our tight-binding model
in giving a satisfactory description of the band structure,
we use thus a more analytical approach to estimate
the ten Slater-Koster parameters, keeping in mind that a better
set of parameters can be achieved by more refined methods.

We first note that, as mentioned above, the simple form
of the Hamiltonian matrix allows
for an analytical solution of the eigenvalues
on all the high-symmetry points $\Gamma$, H, N, P.
In principle, the identification 
of appropriate ten eigenvalues
on these high-symmetry points
with the corresponding DFT energy levels
can provide a way to determine in an unambiguous way
all the ten tight-binding parameters.
However, as remarked in the previous Section,
one can notice that the parameter $W_{sp\sigma}$
and the combination $S_{pp}^-=S_{pp\sigma}-S_{pp\pi}$,
do not appear at the high-symmetry points $\Gamma$, H, N, P.
Further insight can be thus achieved by considering also the point F,
midway between P and H.
Although with relative less symmetry, the analysis of the eigenvalues
in this point is useful because, as shown in Appendix \ref{a:eigen},
they depend in a independent way of $S_{pp\sigma}$, $S_{pp\pi}$,
allowing thus to evaluate them independently.
A further natural way to estimate $W_{sp\sigma}$,
away from the high-symmetry points, is to associate it with
the properties of the Van Hove singularities.
More precisely, we determine $W_{sp\sigma}$ by fixing that the energy
of the upper Van Hove singularity to be the same in the tight-binding model
as in first-principle calculations.
Details of the procedure are reported in Appendix \ref{a:param}.
The set of all the ten tight-binding parameters,
estimated in this way, is listed in Table \ref{t-TBparam} (first column),\cite{notefermilevel}
and the comparison of the resulting band structure
with DFT calculations is shown in Fig. \ref{f-bands}a.
\begin{table}[t]
\begin{tabular}{|c|r|}
\hline
\hline
$\epsilon_{\rm H} $ & -4.34 \\
$\epsilon_{\rm S_s}$ & -14.63 \\
$\epsilon_{\rm S_p}$ & -3.25 \\
$H_{ss\sigma}$ & -2.73 \\
$S_{ss\sigma}$ & 2.31 \\
$S_{pp\sigma}$ & 1.69 \\
$S_{pp\pi}$ & -0.07 \\
$U_{ss\sigma}$ & 2.81 \\
$V_{sp\sigma}$ & 4.65 \\
$W_{sp\sigma}$ & 3.33 \\
\hline\hline
\end{tabular}
\caption{Slater-Koster tight-binding parameters for H$_3$S
at $P=200$ Gpa.
All terms are in units of eV.}
\label{t-TBparam}
\end{table}
As we can see, in spite of the simplicity of the model,
the agreement with first-principle calculations is remarkable.
Most noticeable is the fair reproduction of the local maximum
close to the Fermi level along the line H-N.
This maximum can be shown to be a three-dimensional Van Hove
singularity and, due to its small average effective
mass,\cite{Pickett2}
it is reflected in a kink/peak
in the density of state close to the Fermi energy (Fig. \ref{f-bands}b). 
Also remarkable is the excellent capture of the 
shape of the Fermi surface sheets, as shown in Fig. \ref{f-bands}c-d.
Such agreement is even more remarkable considering that
no information about the Fermi surface properties was employed to
estimate the tight-binding parameters.
In order to stress the crucial role of the $W_{sp\sigma}$ hopping,
we also show, for comparison, in Fig. \ref{f-bandapp}a  the band structure of the tight-binding
model obtained by setting $W_{sp\sigma}=0$, as implicitly
done in the models considered in Refs. \onlinecite{Igor,Pickett2}.
The local maximum along the cut H-N is in this case missing, as
acknowledged in Ref. \onlinecite{Pickett2}.
The corresponding electron density of states and Fermi surface are
shown in Fig. \ref{f-bandapp}b and \ref{f-bandapp}c, respectively.
Both result to be quite different from the realistic DFT ones.
Note in particular how the peak in the density of states close to the Fermi level
is completely washed out for $W_{sp\sigma}=0$.

In addition to the low-energy properties (Fermi surfaces, etc.) it is
also instructive to check the goodness of the TB results
on a wider energy scale. Such analysis is addressed in Appendix
\ref{a:param}, where the fair agreement between the total DOSs
(as well as the partial DOSs projected on single orbitals)
evaluated in TB and DFT provides further evidence oabout the
quality of the TB description.

\section{Hopping-resolved electron-phonon coupling}

The tight-binding Hamiltonian in Eqs. (\ref{hmatr})-(\ref{vmatr}),
along with the set of Slater-Koster parameters in Table \ref{t-TBparam},
provides a suitable description of the electronic band structure
in the periodic lattice structure.
On the other hand, atomic lattice fluctuations (phonons) from the
equilibrium positions are expected to be strong in such hydrogen-based
compounds, and the coupling between the electronic degrees of freedom
and the lattice distortions is claimed, according a general consensus,
to be the likely pairing mechanism for superconductivity.
Several first-principle studies have quantified this issue
in terms of the electron-phonon Eliashberg function
$\alpha^2F(\omega)$, and of the corresponding
dimensionless electron-phonon coupling constant
$\lambda=2\int_0^\infty
\alpha^2F(\omega)/\omega$.\cite{Igor,Mauri,Duan1,Gross,Pickett,Pickett2,Akashi2015,boeri,ge,li} 
Numerical DFT estimates predict $\lambda \approx 1.5-2$,
signalizing indeed a strong electron-phonon coupling.
An interesting decrease of $\lambda$ as a function of the pressure,
in the high-pressure regime, has also been
predicted.\cite{Mauri,Pickett,Akashi2015} 
It should be keep in mind however that such DFT estimates
are obtained by means of linear response calculations
(with possible inclusions of anharmonic effects)\cite{Mauri}
in a homogeneous system, whereas at such high pressures
the lattice structure of realistic materials is expected
to be closer to a amorphous phase than to a crystal one.
It would be thus high desirable to describe the electron-lattice
interaction at a {\em local} level, in order to include inhomogeneity
in a simple way.

This is indeed the main motivation of the use
of a Slater-Koster formalism in our analysis.
Within the Slater-Koster context,\cite{S-K} the inter-atomic hopping  in
Eqs. (\ref{hmatr})-(\ref{vmatr}) are described in terms
of few two-body energy integrals (Table \ref{t-TBparam})
whose magnitude depends essentially only on the
interatomic distances $R$.
The electronic structure can be thus computed
even in the presence of local inhomogeneous lattice displacements
once the dependence of the Slater-Koster parameters between
two atoms on the corresponding inter-atomic distance is provided.
The essential ingredient is here
the hopping-resolved electron-lattice coupling
$\gamma_\alpha=dt_\alpha/dR_\alpha$,
where $t_\alpha=H_{ss\sigma}$, $S_{ss\sigma}$, $S_{pp\sigma}$, $S_{pp\pi}$,
$U_{ss\sigma}$, $V_{sp\sigma}$, $W_{sp\sigma}$.

\begin{figure}[h!]
\begin{center}
\includegraphics[angle=0, width=7cm,clip=]{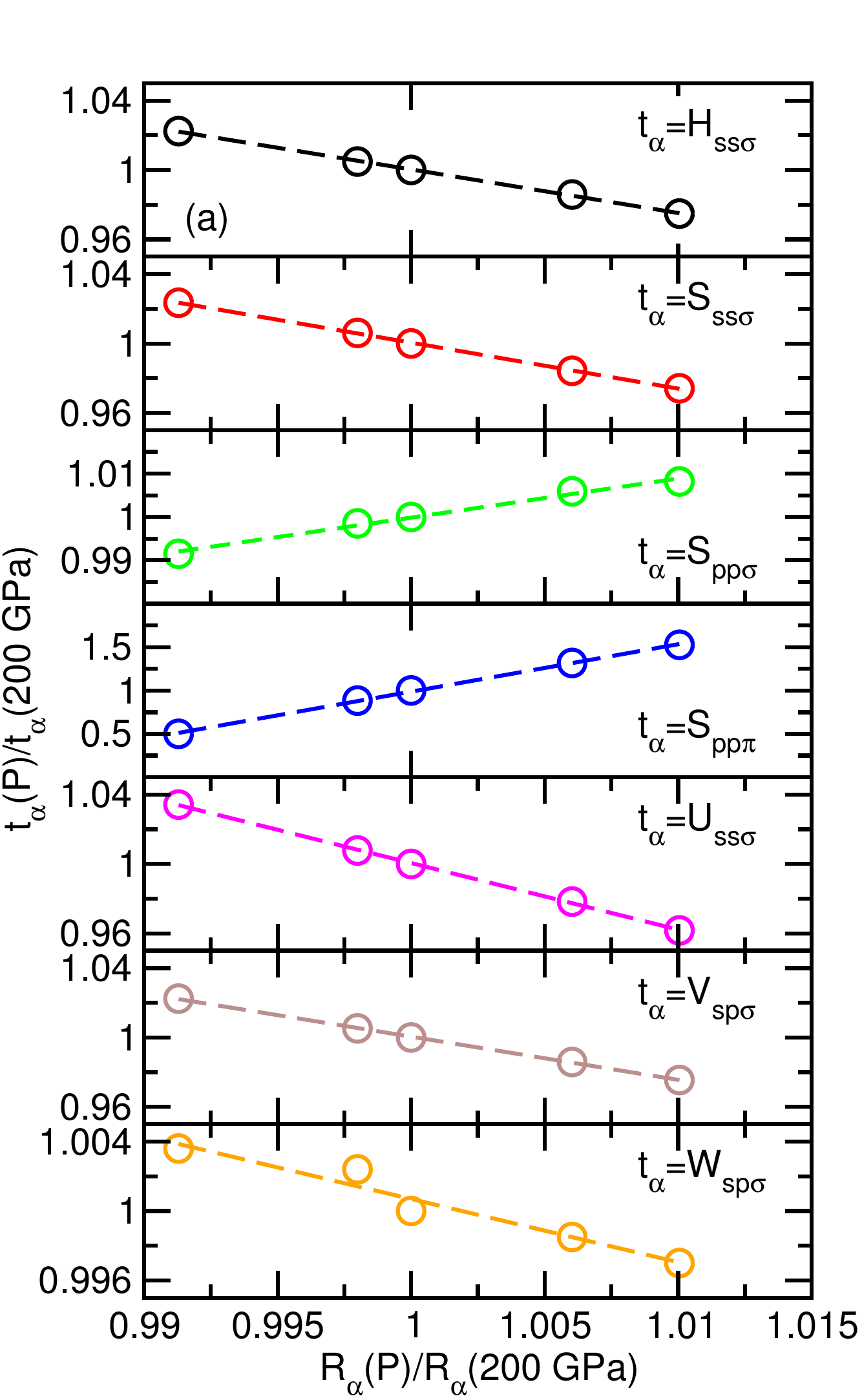}
\includegraphics[angle=0, width=7cm,clip=]{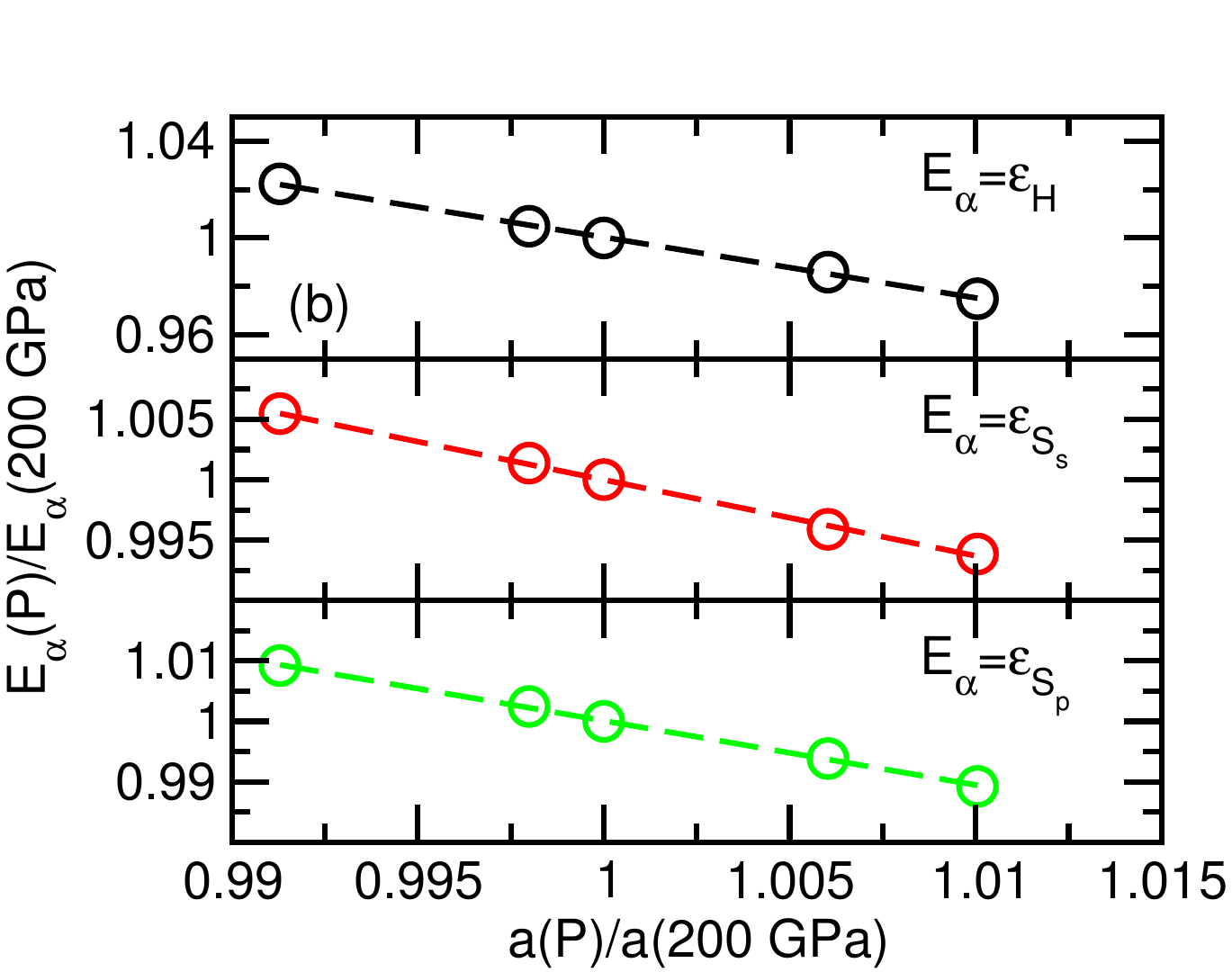}
\caption{(Color online) (a) Dependence of the Slater-Koster hopping
parameters $t_\alpha$ on the relative interatomic distance.
as extracted from the comparison with DFT calculations.
Dashed lines represent linear fits.
(b) Dependence of the on-site
orbital energies $E_\alpha$ on half of the lattice constant $a$.
Open symbols are numerical values
extracted from the comparison with DFT band calculations,
while dashed lines represent linear fits.
}
\label{f-ep1}
\end{center}
\end{figure}

The compelling estimation of the hopping parameters in our model
from the comparison with the first-principle band structure,
and the fair agreement with it, permits us to obtain thus
the hopping-resolved electron-lattice coupling constants
$\gamma_\alpha$
in a controlled way.
More specifically, we apply a isotropic
expansion/contraction of the lattice constant, $2.959$ \AA\, $< d=2a < 3.015$ \AA,
corresponding to the pressure range $180$ GPa $< P < 220$ GPa
where the system has been predicted to be
in the {\em Im}$\bar{3}${\em m} structure.
The same procedure above described
allows us to extract thus the dependence of
the hopping  tight-binding parameters $t_\alpha$
on the relative interatomic distance $R_\alpha$.
The dependence of the parameters $t_\alpha(R_\alpha)$
on the relative atomic distance is shown in Fig. \ref{f-ep1}a.
The dependence of the on-site
orbital energies $E_\alpha=\epsilon_{\rm H}$, $\epsilon_{\rm S_s}$,
$\epsilon_{\rm S_p}$ on the lattice constant is also obtained in this way.
Note however that, within the spirit of a Slater-Koster scheme,
the hopping tight-binding parameters
$t_\alpha$ are thought to depend only on the two-body interatomic distance,
independently of the specific lattice deformation,
and thus to be representative also of phonon lattice displacements,
whereas the modification of the  on-site orbital energies $E_\alpha$ depends strictly on the specific lattice deformation.
The dependence of the on-site orbital energies $E_\alpha$ on the lattice constant, for {\em such} isotropic expansion/contraction, is thus also shown in Fig. \ref{f-ep1}b.
All the tight-binding parameters depends in a linear way on the
variation of the interatomic distance (or of the lattice constant).
The numerical values of the electron-lattice coupling constants
$\gamma_\alpha=dt_\alpha/d R_\alpha$ and
$\chi_\alpha=dE_\alpha/da$, obtained by a linear fit,
is listed in Table \ref{t-elph}.
\begin{table}
\begin{tabular}{|c|r|l|}
\hline
\hline
$\chi_{\epsilon_{\rm H}} $ & 5.49& - \\
$\chi_{\epsilon_{\rm S_s}}$ & 3.09 & -\\
$\chi_{\epsilon_{\rm S_p}}$ & 1.16 & -\\
$\gamma_{H_{ss\sigma}}$ & 2.30 & $R_\alpha=a$\\
$\gamma_{S_{ss\sigma}}$ & -1.18 & $R_\alpha=\sqrt{3}a$\\
$\gamma_{S_{pp\sigma}}$ & 0.29 & $R_\alpha=\sqrt{3}a$\\
$\gamma_{S_{pp\pi}}$ & -0.75 & $R_\alpha=\sqrt{3}a$\\
$\gamma_{U_{ss\sigma}}$ & -3.61 & $R_\alpha=a$\\
$\gamma_{V_{sp\sigma}}$ & -3.87 & $R_\alpha=a$\\
$\gamma_{W_{sp\sigma}}$ & -0.23 & $R_\alpha=\sqrt{3}a$ \\
\hline\hline
\end{tabular}
\caption{First column: Electron-lattice coupling constant $\chi_\alpha$
and $\gamma_\alpha$ of the Slater-Koster tight-binding model
obtained by hydrostatic expansion/contraction.
The values are in units of eV/\AA.
Second column: two-center interatomic distance relative to
each hopping parameter $t_\alpha$.}
\label{t-elph}
\end{table}
It is worth to remark that the absolute value of almost all the
hopping parameters, $|t_\alpha|$, results to decrease with increasing
the interatomic distance, apart from the quantities $S_{pp\sigma}$, $S_{pp\pi}$,
ruling the interatomic hopping between S 3$p$ orbitals, which appear to {\em increase} for larger
interatomic distance.
This result is at apparent odds with the simple Slater-Koster picture
relating the interatomic hopping with two-center orbital integrals.
However it should be kept in mind that, in the mapping with a DFT
calculation, the extracted Slater-Koster parameters represent
{\em effective} quantities that take into account many-body physics.
In this respect, a simple explanation for this unconventional behavior
could be that it results from a {\em reduction} of the underlying microscopic
 two-center orbital integrals compensated, and even overwhelmed,
for the 3$p$-3$p$ hopping, by a {\em reduction} of the screening
effects.
More specific analysis is however needed to investigate in more details,
and possibly confirm, this hypothesis.

\section{Conclusions}

In conclusion, in this paper we have introduced a accurate
tight-binding model, based on the Slater-Koster formalism,
for the electronic band dispersion of H$_3$ at high-pressure
in the {\em Im}$\bar{3}${\em m} structure.
The tight-binding parameters have been numerically evaluated by means
of a direct comparison with first-principle calculations.
We show that the inter-atomic hopping between the 3$s$ and 3$p$ orbitals of
sulphur plays a crucial role in determining the existence and energies
of the Van Hove singularities that hive rise to the peak of density of
states at the Fermi level.
Our tight-binding model provides not only an accurate description
of the band-structure in the whole Brillouin zone, but also an
exceptionally good description of the relevant Fermi surface.
The model here presented provides thus
a fundamental base for the future inclusion of many-body 
(electron-electron, electron-phonon) effects.

In addition,
by computing the effective dependence of the tight-binding parameters
on the interatomic distance, we have also numerically estimated
a hopping-resolved linear electron-lattice coupling, which can be used
for further refinements of the electronic properties in the presence
of lattice disorder.
Furthermore, the knowledge of such inter-atomic electron-lattice couplings
permits in principle to obtain an estimate of the electron-phonon
coupling also in the homogeneous case.
Such modelling, although cannot replace accurate DFT calculations
on a quantitative level, can be nevertheless useful 
in providing a suitable tool for understanding
in a semi-analytical way the role of each microscopical orbital
hybridization (e.g. in the pressure dependence of
$lambda$ in the high-$P$ regime)
and for making qualitative predictions
in more complex and realistic conditions, e.g. in the
presence of disorder etc...
Such computational analyses are however out of the purposes
of the present paper, and they will be addressed in a further work.

\begin{acknowledgements}
We thank M. Eremets, G. Chiarotti, F. Mauri, N. Saini and A. Bianconi
for interesting discussions.
EC acknowledges support from the European project
FP7-PEOPLE-2013-CIG ``LSIE 2D'', Italian National MIUR Prin
project 20105ZZTSE, and the Italian MIUR program ``Progetto
Premiale 2012'' Project ABNANOTECH.
L.O acknowledges CINECA computational resources through the ISCRA B Project n.~IsC24\_mandra03.

\end{acknowledgements}

\begin{appendix}

\section{Analytical insight about the Hamiltonian
at the high-symmetry points}
\label{a:eigen}

In Eqs. (\ref{hmatr})- (\ref{vmatr}) we provided the matricial
expression of the tight-binding Hamiltonian for a generic
momentum ${\bf k}$ in the Brillouin zone.
In this Appendix, for sake of convenience, we summarize
the analytical expression of the eigenvalues obtained
at the high-symmetry points.

\subsection{$\Gamma$ point}

In the basis (\ref{basis}),
the Hamiltonian acquires at the $\Gamma=(0,0,0)$ point
the simple expression:
\begin{eqnarray}
\hat{H}(\Gamma)
&=&
\left(
\begin{array}{cc}
\hat{H}_{\rm HS}(\Gamma) & 0 \\
0 & \hat{H}_{\rm SS}(\Gamma)
\end{array}
\right),
\label{hmatrG}
\end{eqnarray}
where
\begin{eqnarray}
\hat{H}_{\rm HS}(\Gamma)
&=&
\left(
\begin{array}{cccc}
\epsilon_{\rm H} & 2 H_{ss\sigma}  & 2 H_{ss\sigma}  & 2U_{ss\sigma}\\
2 H_{ss\sigma}  & \epsilon_{\rm H} & 2 H_{ss\sigma}  & 2U_{ss\sigma}\\
2 H_{ss\sigma}  & 2 H_{ss\sigma} & \epsilon_{\rm H}  & 2U_{ss\sigma}\\
2U_{ss\sigma} & 2U_{ss\sigma} & 2U_{ss\sigma} & \epsilon_{\rm S_s}+ 8S_{ss\sigma}
\end{array}
\right),
\label{HSmatrG}
\nonumber
\end{eqnarray}
\begin{equation}
\end{equation}
\begin{eqnarray}
\hat{H}_{\rm SS}(\Gamma)
&=&
\left[
\epsilon_{\rm S_p}+\frac{8}{3}S_{pp}^+
\right]
\hat{I}_{3 \times 3}.
\label{SSmatrG}
\end{eqnarray}

Eq. (\ref{SSmatrG}) predicts three degenerate eigenvalues
\begin{eqnarray}
E_{\rm S_p}(\Gamma)
&=&
\epsilon_{\rm S_p}+(8/3)S_{pp}^+,
\end{eqnarray}
with pure character S
$3p_x$, $3p_y$, $3p_z$.

On the other hand,
in order to  deal with (\ref{HSmatrG}),
a better Hilbert basis is:
\begin{eqnarray}
\phi_{i,\Gamma}^\dagger
&=&
(
h_{i,\Gamma_1}^\dagger,
h_{i,\Gamma_2}^\dagger,
h_{i,\Gamma_3}^\dagger,
s_{i}^\dagger,
p_{i,x}^\dagger,
p_{i,y}^\dagger,
p_{i,z}^\dagger
),
\label{basisG}
\end{eqnarray}
where
\begin{eqnarray}
h_{i,\Gamma_1}^\dagger
&=&
\frac{1}{\sqrt{6}}
\left(
h_{i,1}^\dagger
+
h_{i,2}^\dagger
-
2h_{i,3}^\dagger,
\right),
\end{eqnarray}
\begin{eqnarray}
h_{i,\Gamma_2}^\dagger
&=&
\frac{1}{\sqrt{2}}
\left(
h_{i,1}^\dagger
-
h_{i,2}^\dagger,
\right),
\end{eqnarray}
\begin{eqnarray}
h_{i,\Gamma_3}^\dagger
&=&
\frac{1}{\sqrt{3}}
\left(
h_{i,1}^\dagger
+
h_{i,2}^\dagger
+
h_{i,3}^\dagger,
\right).
\end{eqnarray}

In this basis $\hat{H}_{\rm HS}(\Gamma)$ reads:
\begin{eqnarray}
\hat{H}_{\rm HS}(\Gamma)
&=&
\left(
\begin{array}{cccc}
H_{\rm \Gamma_1}(\Gamma) & 0 & 0 & 0\\
0 & H_{\rm \Gamma_2}(\Gamma) & 0 & 0\\
0 & 0 & H_{\rm \Gamma_3}(\Gamma) & 2\sqrt{3}U_{ss\sigma} \\
0 & 0 & 2\sqrt{3}U_{ss\sigma} & S_{\rm S_s}(\Gamma)
\end{array}
\right),
\label{HSmatrG3}
\nonumber
\end{eqnarray}
\begin{equation}
\end{equation}
where
\begin{eqnarray}
H_{\rm \Gamma_1}(\Gamma)
&=&
H_{\rm \Gamma_2}(\Gamma)
=
\epsilon_{\rm H} -2 H_{ss\sigma},
\end{eqnarray}
\begin{eqnarray}
H_{\rm \Gamma_3}(\Gamma)
&=&
\epsilon_{\rm H} +4 H_{ss\sigma},
\end{eqnarray}
\begin{eqnarray}
S_{\rm S_s}(\Gamma)
&=&
\epsilon_{\rm S_s}+8S_{ss\sigma}.
\end{eqnarray}

We get thus an analytical insight
on the further eigenvalues. In particular
Eq. (\ref{HSmatrG3})
 predicts
two degenerate eigenvalues,
\begin{eqnarray}
E_{\rm H}(\Gamma)
&=&
\epsilon_{\rm H} -2 H_{ss\sigma},
\end{eqnarray}
with pure H character,
and a bonding/antibonding couple of eigenvalues:
\begin{eqnarray}
E_{{\rm H-S},\pm}(\Gamma)
&=&
\frac{1}{2}
\left[
\epsilon_{\rm H}+\epsilon_{\rm S_s}
+4H_{ss\sigma}+8S_{ss\sigma}
\right.
\nonumber\\
&&
\pm
\left.
\sqrt{(\epsilon_{\rm H}-\epsilon_{\rm S_s}+4H_{ss\sigma}-8S_{ss\sigma})^2+48U_{ss\sigma}^2}
\right],
\nonumber
\end{eqnarray}
\begin{equation}
\end{equation}
resulting from
the hybridization of the H $1s$ with the S $3s$ orbitals.

\subsection{H point}

The Hamiltonian acquires a simple and analytical
expression also at the H$=(\pi/a,0,0)$ point.
In this case we have:
\begin{eqnarray}
\hat{H}({\rm H})
&=&
\left(
\begin{array}{cc}
\hat{H}_{\rm HS}({\rm H}) & 0 \\
0 & \hat{H}_{\rm SS}({\rm H})
\end{array}
\right),
\label{hmatrH}
\end{eqnarray}
where
\begin{eqnarray}
\hat{H}_{\rm HS}({\rm H})
&=&
\left(
\begin{array}{cccc}
\epsilon_{\rm H} & 2 H_{ss\sigma}  & 2 H_{ss\sigma}  & -2U_{ss\sigma}\\
2 H_{ss\sigma}  & \epsilon_{\rm H} & -2 H_{ss\sigma}  & 2U_{ss\sigma}\\
2 H_{ss\sigma}  & -2 H_{ss\sigma} & \epsilon_{\rm H}  & 2U_{ss\sigma}\\
-2U_{ss\sigma} & 2U_{ss\sigma} & 2U_{ss\sigma} & \epsilon_{\rm S_s}- 8S_{ss\sigma}
\end{array}
\right),
\label{HSmatrH}
\nonumber
\end{eqnarray}
\begin{equation}
\end{equation}
and
\begin{eqnarray}
\hat{H}_{\rm SS}({\rm H})
&=&
\left[
\epsilon_{\rm S_p}-\frac{8}{3}S_{pp}^+
\right]
\hat{I}_{3 \times 3}.
\label{SSmatrH}
\end{eqnarray}

Eq. (\ref{hmatrH}) has the same formal structure as
(\ref{hmatrG}), with some sign replacements.

The Hamiltonian sub-block $\hat{H}_{\rm SS}({\rm H})$ gives
rise to three degenerate eigenvalues with pure character S
$3p$ character:
\begin{eqnarray}
E_{\rm S_p}({\rm H})
&=&
\epsilon_{\rm S_p}-(8/3)S_{pp}^+.
\end{eqnarray}

On the other hand,
a better appropriate basis can be found also in this case,
in order to deal with the sub-block
$\hat{H}_{\rm HS}({\rm H})$.
We write thus:
\begin{eqnarray}
\phi_{i,H}^\dagger
&=&
(
h_{i,H_1}^\dagger,
h_{i,H_2}^\dagger,
h_{i,H_3}^\dagger,
s_{i}^\dagger,
p_{i,x}^\dagger,
p_{i,y}^\dagger,
p_{i,z}^\dagger
),
\label{basisH}
\end{eqnarray}
where
\begin{eqnarray}
h_{i,H_1}^\dagger
&=&
\frac{1}{\sqrt{6}}
\left(
h_{i,1}^\dagger
-
h_{i,2}^\dagger
+
2h_{i,3}^\dagger,
\right),
\end{eqnarray}
\begin{eqnarray}
h_{i,H_2}^\dagger
&=&
\frac{1}{\sqrt{2}}
\left(
h_{i,1}^\dagger
+
h_{i,2}^\dagger,
\right),
\end{eqnarray}
\begin{eqnarray}
h_{i,H_3}^\dagger
&=&
\frac{1}{\sqrt{3}}
\left(
h_{i,1}^\dagger
-
h_{i,2}^\dagger
-
h_{i,3}^\dagger,
\right).
\end{eqnarray}

In this basis $\hat{H}_{\rm HS}({\rm H})$ acquires the block form:
\begin{eqnarray}
\hat{H}_{\rm HS}({\rm H})
&=&
\left(
\begin{array}{cccc}
H_{\rm H_1}({\rm H}) & 0 & 0 & 0\\
0 & H_{\rm H_2}({\rm H}) & 0 & 0\\
0 & 0 & H_{\rm H_3}({\rm H}) & -2\sqrt{3}U_{ss\sigma} \\
0 & 0 & -2\sqrt{3}U_{ss\sigma} & S_{\rm S_s}({\rm H})
\end{array}
\right),
\label{HSmatrH3}
\nonumber
\end{eqnarray}
\begin{equation}
\end{equation}
where
\begin{eqnarray}
H_{\rm H_1}({\rm H})
&=&
H_{\rm H_2}({\rm H})
=
\epsilon_{\rm H} +2 H_{ss\sigma},
\end{eqnarray}
\begin{eqnarray}
H_{\rm H_3}({\rm H})
&=&
\epsilon_{\rm H} -4 H_{ss\sigma},
\end{eqnarray}
\begin{eqnarray}
S_{\rm S_s}({\rm H})
&=&
\epsilon_{\rm S_s}-8S_{ss\sigma}.
\end{eqnarray}

Similar as at the $\Gamma$ point, we have thus
two degenerate eigenvalues with pure H character,
\begin{eqnarray}
E_{\rm H}({\rm H})
&=&
\epsilon_{\rm H} +2 H_{ss\sigma},
\end{eqnarray}
and a bonding/antibonding couple of eigenvalues
coming from
the hybridization of the H $1s$ with the S $3s$ orbitals:
\begin{eqnarray}
E_{{\rm H-S},\pm}({\rm H})
&=&
\frac{1}{2}
\left[
\epsilon_{\rm H}+\epsilon_{\rm S_s}
-4H_{ss\sigma}-8S_{ss\sigma}
\right.
\nonumber\\
&&
\pm
\left.
\sqrt{(\epsilon_{\rm H}-\epsilon_{\rm S_s}-4H_{ss\sigma}+8S_{ss\sigma})^2+48U_{ss\sigma}^2}
\right].
\nonumber
\end{eqnarray}
\begin{equation}
\end{equation}

\subsection{N point}

The single elements (\ref{hsmatr})-(\ref{vmatr}) of the Hamiltonian
(\ref{hmatr})
read at the N$=(\pi/2a,\pi/2a,0)$ point:
\begin{eqnarray}
\hat{H}_\sigma({\rm N})
&=&
\left(
\begin{array}{ccc}
\epsilon_{\rm H} & 2 H_{ss\sigma}  & 0 \\
2 H_{ss\sigma}  & \epsilon_{\rm H} & 0 \\
0 & 0 & \epsilon_{\rm H}
\end{array}
\right),
\end{eqnarray}
\begin{eqnarray}
\hat{S}_\sigma({\rm N})
&=&
\epsilon_{\rm S_s},
\end{eqnarray}
\begin{eqnarray}
\hat{S}_\pi({\rm N})
&=&
\epsilon_{\rm S_p}
\hat{I}_{3 \times 3}
+
\nonumber\\
&&
\frac{8}{3}\left(
\begin{array}{ccc}
0
& S_{pp}^-
& 0
\\
S_{pp}^-
& 0
& 0
\\
0
& 0
& 0
\end{array}
\right),
\end{eqnarray}
\begin{eqnarray}
\hat{U}_\sigma({\rm N})
&=&
\left(
\begin{array}{c}
0 \\
0 \\
2U_{ss\sigma}
\end{array}
\right),
\end{eqnarray}
\begin{eqnarray}
\hat{W}_\sigma({\rm N})
&=&
\left(
\begin{array}{ccc}
0 & 0&0
\end{array}
\right),
\end{eqnarray}
\begin{eqnarray}
\hat{V}_\sigma({\rm N})
&=&
\left(
\begin{array}{ccc}
2iV_{sp\sigma} & 0 & 0 \\
0 & 2iV_{sp\sigma} & 0 \\
0 & 0 & 0
\end{array}
\right).
\label{vmatrN}
\end{eqnarray}

A proper basis can be found in this case to be:
$$
\phi_{i,N}^\dagger
=$$
\begin{eqnarray}(
h_{i,N_+}^\dagger,
p_{i,N_+}^\dagger,
h_{i,N_-}^\dagger,
p_{i,N_-}^\dagger,
h_{i,3}^\dagger,
s_{i}^\dagger,
p_{i,z}^\dagger
),
\label{basisN}
\end{eqnarray}
where
$h_{i,N_\pm}^\dagger=[h_{i,1}^\dagger \pm h_{i,2}^\dagger]/\sqrt{2}$
and $p_{i,N_\pm}^\dagger=[p_{i,x}^\dagger \pm
p_{i,y}^\dagger]/\sqrt{2}$.

In this basis the Hamiltonian has the simple block form:
\begin{eqnarray}
\hat{H}({\rm N})
&=&
\left(
\begin{array}{cccc}
\hat{H}_{++}({\rm N}) & 0 & 0 & 0\\
0 & \hat{H}_{--}({\rm N}) & 0 & 0\\
0 & 0 & \hat{H}_{\rm H_3-S_s}({\rm N}) & 0 \\
0 & 0 & 0 &H_{\rm S_p}({\rm N})
\end{array}
\right),
\label{hmatrN}
\nonumber
\end{eqnarray}
\begin{equation}
\end{equation}
where
\begin{eqnarray}
\hat{H}_{++}({\rm N})
&=&
\left(
\begin{array}{cc}
\epsilon_{\rm H} + 2 H_{ss\sigma}  & 2iV_{sp\sigma} \\
-2iV_{sp\sigma}  & \epsilon_{\rm S_p}- S_{pp}^-
\end{array}
\right),
\label{ppmatrN}
\end{eqnarray}
\begin{eqnarray}
\hat{H}_{--}({\rm N})
&=&
\left(
\begin{array}{cc}
\epsilon_{\rm H} - 2 H_{ss\sigma}  & -2iV_{sp\sigma} \\
2iV_{sp\sigma}  & \epsilon_{\rm S_p}+ S_{pp}^-
\end{array}
\right),
\label{mmmatrN}
\end{eqnarray}
\begin{eqnarray}
\hat{H}_{\rm H_3-S_s}({\rm N})
&=&
\left(
\begin{array}{cc}
\epsilon_{\rm H}  & 2U_{ss\sigma} \\
2U_{ss\sigma}  & \epsilon_{\rm S_s}
\end{array}
\right),
\label{3smatrN}
\end{eqnarray}
and
\begin{eqnarray}
H_{\rm S_p}({\rm N})
&=&
\epsilon_{\rm S_p}.
\end{eqnarray}

The $2 \times 2$ blocks $\hat{H}_{++}({\rm N})$, $\hat{H}_{--}({\rm N})$
are associated with the hybridization of he hydrogen atoms 1 and 2
with the S 3$p_x$, 3$p_y$ orbitals,
and they give rise to the eigenvalues:
\begin{eqnarray}
E_{++,\pm}({\rm N})
&=&
\frac{1}{2}
\left[
\epsilon_{\rm H}+\epsilon_{\rm S_p}
+2H_{ss\sigma}-2S_{pp}^-
\right.
\nonumber\\
&&
\pm
\left.
\sqrt{(\epsilon_{\rm H}-\epsilon_{\rm S_p}+2H_{ss\sigma}+2S_{pp}^-)^2+16V_{sp\sigma}^2}
\right],
\nonumber
\end{eqnarray}
\begin{equation}
\end{equation}
\begin{eqnarray}
E_{--,\pm}({\rm N})
&=&
\frac{1}{2}
\left[
\epsilon_{\rm H}+\epsilon_{\rm S_p}
-2H_{ss\sigma}+2S_{pp}^-
\right.
\nonumber\\
&&
\pm
\left.
\sqrt{(\epsilon_{\rm H}-\epsilon_{\rm S_p}-2H_{ss\sigma}-2S_{pp}^-)^2+16V_{sp\sigma}^2}
\right].
\nonumber
\end{eqnarray}
\begin{equation}
\end{equation}
The $2 \times 2$ block $\hat{H}_{\rm H_3-S_s}({\rm N})$ describes
the hybridization between the hydrogen atom 3 with the
3$s$ sulphur orbitals, which gives rise to the bonding/antibonding
energy levels:
\begin{eqnarray}
E_{\rm H_3-S_s,\pm}({\rm N})
&=&
\frac{1}{2}
\left[
\epsilon_{\rm H}+\epsilon_{\rm S_s}
\right.
\nonumber\\
&&
\pm
\left.
\sqrt{(\epsilon_{\rm H}-\epsilon_{\rm S_s})^2+16U_{ss\sigma}^2}
\right].
\end{eqnarray}
Finally,  the term $H_{\rm S_p}({\rm N})$
represents a state with pure 3 $p_z$ sulphur character,
with a straightforward eigenvalue
\begin{eqnarray}
E_{\rm S_p}({\rm N})
&=&
\epsilon_{\rm S_p}.
\end{eqnarray}

\subsection{P point}

The  Hamiltonian acquires a simple analytical expression
also at the P$=(\pi/2a,\pi/2a,\pi/2a)$ point, where
we can write
\begin{eqnarray}
\hat{H}_\sigma({\rm P})
&=&
\epsilon_{\rm H} \hat{I}_{3 \times 3},
\end{eqnarray}
\begin{eqnarray}
\hat{S}_\sigma({\rm P})
&=&
\epsilon_{\rm S_s},
\end{eqnarray}
\begin{eqnarray}
\hat{S}_\pi({\rm P})
&=&
\epsilon_{\rm S_p}
\hat{I}_{3 \times 3},
\end{eqnarray}
\begin{eqnarray}
\hat{U}_\sigma({\rm P})
&=&
\left(
\begin{array}{c}
0 \\
0 \\
0
\end{array}
\right),
\end{eqnarray}
\begin{eqnarray}
\hat{W}_\sigma({\rm P})
&=&
\left(
\begin{array}{ccc}
0 & 0 & 0
\end{array}
\right),
\end{eqnarray}
\begin{eqnarray}
\hat{V}_\sigma({\rm P})
&=&
2iV_{sp\sigma}
\hat{I}_{3 \times 3}.
\end{eqnarray}

In this case, the Hamiltonian acquires a block expression
after a simple re-ordering of the vector (\ref{basis}):
\begin{eqnarray}
\phi_{i,P}^\dagger
&=&
(
h_{i,1}^\dagger,
p_{i,x}^\dagger,
h_{i,2}^\dagger,
p_{i,y}^\dagger,
h_{i,3}^\dagger,
p_{i,z}^\dagger,
s_{i}^\dagger
).
\label{basisP}
\end{eqnarray}
Using this basis, we obtain thus:
\begin{eqnarray}
\hat{H}({\rm P})
&=&
\left(
\begin{array}{cccc}
\hat{H}_{\rm HS}({\rm P}) & 0 & 0 & 0\\
0 & \hat{H}_{\rm HS}({\rm P}) & 0 & 0 \\
0 & 0 & \hat{H}_{\rm HS}({\rm P}) & 0 \\
0 & 0 & 0 & \epsilon_{\rm S_s}
\end{array}
\right),
\label{hmatrP}
\end{eqnarray}
where
\begin{eqnarray}
\hat{H}_{\rm HS}({\rm P})
&=&
\left(
\begin{array}{cc}
\epsilon_{\rm H} & 2iV_{sp\sigma} \\
-2iV_{sp\sigma} & \epsilon_{\rm S_p}
\end{array}
\right).
\label{hsmatrP}
\end{eqnarray}

The energy spectrum is characterized thus by a single energy level with pure S $3s$ character,
\begin{eqnarray}
E_{\rm S_s}({\rm P})
&=&
\epsilon_{\rm S_s},
\end{eqnarray}
plus
three couples of degenerate (bonding/antibonding) eigenvalues
arising from the hybridization of the H $1s$ with the S $3p$ orbitals:
\begin{eqnarray}
E_{{\rm H-S},\pm}({\rm P})
&=&
\frac{1}{2}
\left[
\epsilon_{\rm H}+\epsilon_{\rm S_p}
\right.
\nonumber\\
&&
\pm
\left.
\sqrt{(\epsilon_{\rm H}-\epsilon_{\rm S_p})^2+16V_{sp\sigma}^2}
\right].
\end{eqnarray}

\subsection{F point}
\label{appF}

Along the high-symmetry points $\Gamma$, H, N and P,
it is useful to consider also the Hamiltonian matrix structure
(and corresponding eigenvalues/eigenvectors) at the point
F$=(3\pi/4a,\pi/4a,\pi/4a)$.
The single elements (\ref{hsmatr})-(\ref{vmatr}) of the Hamiltonian
(\ref{hmatr}) acquire a simple form also at
the F$=(3\pi/4a,\pi/4a,\pi/4a)$ point.
At this points the single elements
(\ref{hsmatr})-(\ref{vmatr}) of the Hamiltonian
(\ref{hmatr})
read 
\begin{eqnarray}
\hat{H}_\sigma({\rm F})
&=&
\left(
\begin{array}{ccc}
\epsilon_{\rm H} & \sqrt{2} H_{ss\sigma}  & \sqrt{2} H_{ss\sigma} \\
\sqrt{2} H_{ss\sigma} & \epsilon_{\rm H} & -\sqrt{2} H_{ss\sigma} \\
\sqrt{2} H_{ss\sigma} & -\sqrt{2} H_{ss\sigma} & \epsilon_{\rm H}
\end{array}
\right),
\end{eqnarray}
\begin{eqnarray}
\hat{S}_\sigma({\rm F})
&=&
\epsilon_{\rm S_s}
- 2\sqrt{2} S_{ss\sigma}
\end{eqnarray}
\begin{eqnarray}
\hat{S}_\pi({\rm F})
&=&
\epsilon_{\rm S_p}
\hat{I}_{3\times 3}
+
\nonumber\\
&&
\frac{2\sqrt{2}}{3}\left(
\begin{array}{ccc}
-S_{pp}^+
& S_{pp}^-
& S_{pp}^-
\\
S_{pp}^-
& -S_{pp}^+
& -S_{pp}^-
\\
S_{pp}^-
& -S_{pp}^-
& -S_{pp}^+
\end{array}
\right),
\end{eqnarray}
\begin{eqnarray}
\hat{U}_\sigma({\rm F})
&=&
\sqrt{2}U_{ss\sigma}
\left(
\begin{array}{c}
-1 \\
1 \\
1
\end{array}
\right),
\end{eqnarray}
\begin{eqnarray}
\hat{W}_\sigma({\rm F})
&=&
i\frac{2\sqrt{2}W_{sp\sigma}}{\sqrt{3}}
\left(
\begin{array}{ccc}
1 & -1 & -1
\end{array}
\right),
\end{eqnarray}
\begin{eqnarray}
\hat{V}_\sigma({\rm F})
&=&
i\sqrt{2}V_{sp\sigma}
\hat{I}_{3\times 3}.
\end{eqnarray}

An appropriate base in this case is:
\begin{eqnarray}
\phi_{i,F}^\dagger
&=&
(
h_{i,F_1}^\dagger,
p_{i,F_1}^\dagger,
h_{i,F_2}^\dagger,
p_{i,F_2}^\dagger,
h_{i,F_3}^\dagger,
s_{i}^\dagger,
h_{i,F_3}^\dagger,
p_{i,F_3}^\dagger
),
\label{basisF}
\nonumber
\end{eqnarray}
\begin{equation}
\end{equation}
where
\begin{eqnarray}
h_{i,F_1}^\dagger
&=&
\frac{1}{\sqrt{6}}
\left(
2h_{i,1}^\dagger
+
h_{i,2}^\dagger
+
h_{i,3}^\dagger,
\right),
\end{eqnarray}
\begin{eqnarray}
h_{i,F_2}^\dagger
&=&
\frac{1}{\sqrt{2}}
\left(
h_{i,2}^\dagger
-
h_{i,3}^\dagger,
\right),
\end{eqnarray}
\begin{eqnarray}
h_{i,F_3}^\dagger
&=&
\frac{1}{\sqrt{3}}
\left(
h_{i,1}^\dagger
-
h_{i,2}^\dagger
-
h_{i,3}^\dagger,
\right),
\end{eqnarray}
\begin{eqnarray}
p_{i,F_1}^\dagger
&=&
\frac{1}{\sqrt{6}}
\left(
2p_{i,x}^\dagger
+
p_{i,y}^\dagger
+
p_{i,z}^\dagger,
\right),
\end{eqnarray}
\begin{eqnarray}
p_{i,F_2}^\dagger
&=&
\frac{1}{\sqrt{2}}
\left(
p_{i,y}^\dagger
-
p_{i,z}^\dagger,
\right),
\end{eqnarray}
\begin{eqnarray}
p_{i,F_3}^\dagger
&=&
\frac{1}{\sqrt{3}}
\left(
p_{i,x}^\dagger
-
p_{i,y}^\dagger
-
p_{i,z}^\dagger,
\right).
\end{eqnarray}

In this basis the Hamiltonian takes the simple block expression:
\begin{eqnarray}
\hat{H}({\rm F})
&=&
\left(
\begin{array}{ccc}
\hat{H}_{\rm 2 \times 2}({\rm F}) & 0 & 0 \\
0 & \hat{H}_{\rm 2 \times 2}({\rm F}) & 0  \\
0 & 0 & \hat{H}_{\rm 3 \times 3}({\rm F}) 
\end{array}
\right),
\label{hmatrF}
\end{eqnarray}
where
\begin{eqnarray}
\hat{H}_{\rm 2 \times 2}({\rm F})
&=&
\left(
\begin{array}{cc}
\epsilon_{\rm H}+\sqrt{2}H_{ss\sigma} & \sqrt{2}iV_{sp\sigma} \\
-\sqrt{2}iV_{sp\sigma} & \epsilon_{\rm S_p}-2\sqrt{2}S_{pp\pi}
\end{array}
\right),
\label{hsmatrF22}
\nonumber
\end{eqnarray}
\begin{equation}
\end{equation}
and
\begin{eqnarray}
\hat{H}_{\rm 3 \times 3}({\rm F})
&=&
\left(
\begin{array}{ccc}
\epsilon_{\rm S_s}  & 0 & 0\\
0 & \epsilon_{\rm H} & 0 \\
0 & 0 & \epsilon_{\rm S_p}
\end{array}
\right)
\nonumber\\
&&
+
\left(
\begin{array}{ccc}
- 2\sqrt{2} S_{ss\sigma} & -\sqrt{6} U_{ss\sigma} & i2\sqrt{2}W_{sp\sigma}\\
-\sqrt{6} U_{ss\sigma}  & -2\sqrt{2}H_{ss\sigma} & \sqrt{2}iV_{sp\sigma} \\
-i2\sqrt{2}W_{sp\sigma} & -\sqrt{2}iV_{sp\sigma} & -2\sqrt{2}S_{pp\sigma}
\end{array}
\right).
\label{hsmatrF33}
\nonumber
\end{eqnarray}
\begin{equation}
\label{eq.71}
\end{equation}
The degenerate blocks $\hat{H}_{\rm 2 \times 2}({\rm F})$ gives rise
to (double degenerate) bonding/antibonding levels
with mixed character (H 1$s$-S 3$p$) and
energies
\begin{eqnarray}
&&E_{{\rm 2\times 2},\pm}({\rm F})
=
\frac{1}{2}
\left[
\epsilon_{\rm H}+ \epsilon_{\rm S_p}+\sqrt{2}H_{ss\sigma} -2\sqrt{2}S_{pp\pi}
\right.
\nonumber\\
&&
\pm
\left.
\sqrt{(
\epsilon_{\rm H}- \epsilon_{\rm S_p}+\sqrt{2}H_{ss\sigma} +2\sqrt{2}S_{pp\pi}
)^2
+8V_{sp\sigma}^2}
\right].
\nonumber
\end{eqnarray}
\begin{equation}
\label{eq.72}
\end{equation}
On the other hand, the $3 \times 3$ block
$\hat{H}_{\rm 3 \times 3}({\rm F})$ does not allow
for a simple analytical solution of the eigenvalues,
but the orbital character of the eigenvectors can be predicted
to be a mix of all the three orbital species, (H 1$s$ +S 3$s$ + S 3$p$).

\section{Slater-Koster parameters of the tight-binding model}
\label{a:param}

In this Appendix we discuss the method we used to estimate the ten
tight-binding parameters, including the orbital energy potentials
and the hopping parameters described within the
Slater-Koster formalism.

We notice that, as discussed in the previous Appendix, the Hamiltonian (\ref{vmatr})
allows for an analytical solution of the eigenvalues in all the main
high-symmetry points $\Gamma$, H, N, P, of the Brillouin zone.
By a closer inspection of the atomic/orbital character,
it is possible to identify almost each eigenvalue with a corresponding
energy level in the DFT band structure.

\begin{table}
\begin{tabular}{|c|c|c|c|c|c|c|}
\hline
\hline
DFT  & H $1s$ & S $3s$ & S $3p$ & IS & other & TB \\
\hline\hline
\multicolumn{7}{|c|}{$\Gamma$ point} \\
\hline
-25.23         & 15 \%   & 34\%     &  -         & 51 \%  & $< 1\%$ & $E_{{\rm H-S},-}(\Gamma)$       \\
0.88$^\S$     &    -   & -           &  68\%   & 29 \%   & 3 \%& $E_{\rm S_p}(\Gamma)$ (*) \\
1.13$^\dagger$     & 49\%    & -           &  -         & 35 \% & 16 \%   & $E_{\rm H}(\Gamma)$ (*)\\
7.93            & 25 \%    & 51 \%   & -          & 23\%   & 1 \%& $E_{{\rm H-S},+}(\Gamma)$ (*) \\
\hline\hline
\multicolumn{7}{|c|}{H point} \\
\hline
-12.81         & 10 \%   & 67 \%  & -       & 23 \% & $< 1\%$  & $E_{{\rm H-S},-}({\rm H})$    \\
-9.80$^\dagger$   & 36 \%   & -        & -       & 52\%   & 12 \%  & $E_{\rm H}({\rm H})$  (*) \\
-8.25$^\S$   & -         & -         & 43\% & 53 \% & 4\%   & $E_{\rm S_p}({\rm H})$ \\
17.56           & 62\%  & 20\%  & -        & 18 \%  & $< 1\%$  & $E_{{\rm  H-S},+}({\rm H})$ \\
\hline\hline
\multicolumn{7}{|c|}{N point} \\
\hline
-17.82  & 9 \% & 46 \% &  -        & 42 \% & 3\%  & $E_{\rm H_3-S_s,-}({\rm N})$\\
-16.89      & 20 \% & -        &  24\%  & 55\%  & 1 \% & $E_{\alpha\alpha,\pm}({\rm N})^\P$ \\
-5.05        & 28 \% & -        & 35\%   & 36\%  & 1 \%&  $E_{\alpha\alpha,\pm}({\rm N})^\P$\\
-3.25    & -        & -        & 56\%   & 41 \% & 3 \% & $E_{\rm S_p}({\rm N})$ (*)\\
-1.86    & 23\%  & 21\%  &   -       & 41 \% & 15\% & $E_{\rm H_3-S_s,+}({\rm N})$ (*)\\
12.75        & 29 \% & -        & 31\%   & 36 \% & 4\% &  $E_{\alpha\alpha,\pm}({\rm N})^\P$\\
12.77        & 15 \% & -        & 40\%   & 39 \% & 6 \% & $E_{\alpha\alpha,\pm}({\rm N})^\P$\\
\hline\hline
\multicolumn{7}{|c|}{P point} \\
\hline
-14.63          & -        & 65 \%  & -         & 33 \% & 2\% & $E_{\rm S_s}({\rm P})$ (*)\\
-13.11$^\S$  &  21 \% & -        & 24\%   & 53 \% & 2\% & $E_{{\rm H-S},-}({\rm P})$ (*)\\
12.99$^\S$    & 3 \%    & -        & 45\%   & 32 \% & 20 \% & $E_{{\rm H-S},+}({\rm P})$ \\ 
\hline\hline
\multicolumn{7}{|c|}{F point} \\
\hline
-15.45          & 6 \% & 57 \%  & $< 1\%$         & 35 \% & 2\% & $E_{ 3 \times 3}({\rm F})^\P$ \\
-13.99$^\dagger$      &  21 \% & -        & 22\%   & 54 \% & 3\% & $E_{2\times 2,-}({\rm F})$ \\
-2.98          & 12  \%  & $< 1\%$ & 52 \% & 33 \% & 3\% & $E_{3\times 3}({\rm F})^\P$ \\
1.44$^\dagger$      &  15 \% & -        & 21\%   & 47 \% & 17\% & $E_{2\times 2,+}({\rm F})$ (*)\\
4.83    & 4 \%    & 14 \%       & $< 1\%$  & 57 \% & 25 \% & $E_{3\times 3}({\rm F})^\P$ \\ 
\hline\hline
\end{tabular}
\caption{First principle atomic/orbital character for the relevant energy
levels at the main high symmetry points $\Gamma$, H, N, P, F of the
Brillouin zone. The
content of the orbitals H $1s$, S $3s$ and S $3p$ is explicitly
reported,
as well as the contribution from the interstitial states (IS) and from
other orbitals (``other''). The right column provides the identification of each
DFT level
with the corresponding eigenvalues of the tight-binding model.
For the levels marked by $\P$ at the N point, it is not possible
a straightforward identification with the each of the four eigenstates
$E_{\alpha\alpha,\pm}({\rm N})=E_{++,\pm}({\rm N}),E_{--,\pm}({\rm N})$.
In similar way, it is not possible associate straightforwardly the levels marked with 
$\P$ at the F point with the three eigenvalues of the block $\hat{H}_{\rm 3 \times 3}({\rm F})$.
Energy levels marked with (*) represent energy levels used
to determine the tight-binding parameters.
\newline
$^\dagger$ double degenerate level\newline
$^\S$ triple degenerate level\newline
}
\label{t-dft}
\end{table}

In Table \ref{t-dft} we report the DFT atomic/orbital content 
for each relevant energy level at the high-symmetry points
$\Gamma$, H, N, P.
The corresponding eigenvalue in the tight-binding formalism,
obtained by comparing the atomic/orbital content,
is also shown.
The identification of each DFT energy level with a corresponding
tight-binding level, along with the analytical expression
of the tight-binding eigenvalues in terms of the Slater-Koster
parameters, allows in principle for a closed set of equations
that can be inverted to obtained the microscopic tight-binding
parameter.
This procedure however can be used only at a limited extend.
On one hand, four of the DFT non degenerate levels at the N point
present the same mix of orbital content, so that it results impossible
to identify each of them
with a single tight-binding level belonging to the subspace $E_{++,\pm}({\rm N})$, $E_{--,\pm}({\rm N})$.
Even excluding from the analysis these levels, the DFT-TB one-to-one mapping still provides
in principle 14 equations, more than the number of
parameters to be evaluated. 
The system appears thus {\em overdetermined}.
On the other hand, as evident from Table \ref{t-dft}, the Hilbert set
of atomic orbitals
here considered does not encompass the total orbital weight,
with a sizable contribution from S $3d$, H $1p$ being present
(included in the six column).

All these considerations point out the impossibility for the
tight-binding model to reproduce {\em all} the energy levels on
{\em all} the high-symmetry points.
Nevertheless,  a mathematically compelling procedure can be still
defined by requiring that the tight-binding model correctly reproduce
the DFT results on a {\em subset of energy level}.
Clearly, the choice of the selected energy levels is subjective and
its reliability must be checked from the overall agreement of the
whole band structure. 
After a careful analysis, we found the best agreement by using the
DFT-TB levels marked by (*)
at the high-symmetry points $\Gamma$, H, N, P
in Table \ref{t-dft}.
Note that only 8 energy levels are thus selected, determining
the 7 tight-binding parameters 
$\epsilon_{\rm H}$, $\epsilon_{\rm S_s}$, $\epsilon_{\rm S_p}$,
$H_{ss\sigma}$, $S_{ss\sigma}$, 
$U_{ss\sigma}$, $V_{sp\sigma}$, 
and the linear combination
$S_{pp}^+=S_{pp\sigma}+2S_{pp\pi}$.
This is due to the fact that the parameter $W_{sp\sigma}$
does not appear in the determination of the tight-binding energy levels
at any high-symmetry point.
In similar way, excluding the levels $E_{++,\pm}({\rm N})$, $E_{--,\pm}({\rm N})$,
also the linear combination $S_{pp}^-=S_{pp\sigma}-S_{pp\pi}$
never appears among the eigenvalues of the tight-binding model
on any high-symmetry points $\Gamma$-H-N-P.

The remaining quantities $W_{sp\sigma}$, $S_{pp}^-$ must be
determined thus by imposing a DFT-TB correspondence on some
band structure property not related to the high-symmetry point.

A step forwards in determine these parameters comes from considering
also the lower-symmetry point F$=(3\pi/4a,\pi/4a,\pi/4a)$.
As shown in Section \ref{appF}[Eqs. (\ref{hsmatrF22})-(\ref{eq.72})], this results to be very profitable because
on this point the Slater-Koster parameters $S_{pp\sigma}$ and $S_{pp\pi}$
appear in an independent way in different eigenvalues.
Quite useful appears in particular the level $E_{2\times 2,+}(F)$,
whose energy is ruled by only $S_{pp\pi}$ (and not by $S_{pp\sigma}$),
and which, after an identification with the DFT energy levels by means
of the orbital character (see Table \ref{t-dft}), results to be at $1.44$ eV, relatively close
to the Fermi level.
Including this level among the analytical constraints,
we can therefore estimate, from the DFT calculations
on the $\Gamma$, H, N, P, F points, all the
Slater-Koster tight-binding parameters but $W_{sp\sigma}$.

\begin{figure}[t!]
\begin{center}
\includegraphics[angle=0, width=8cm,clip=]{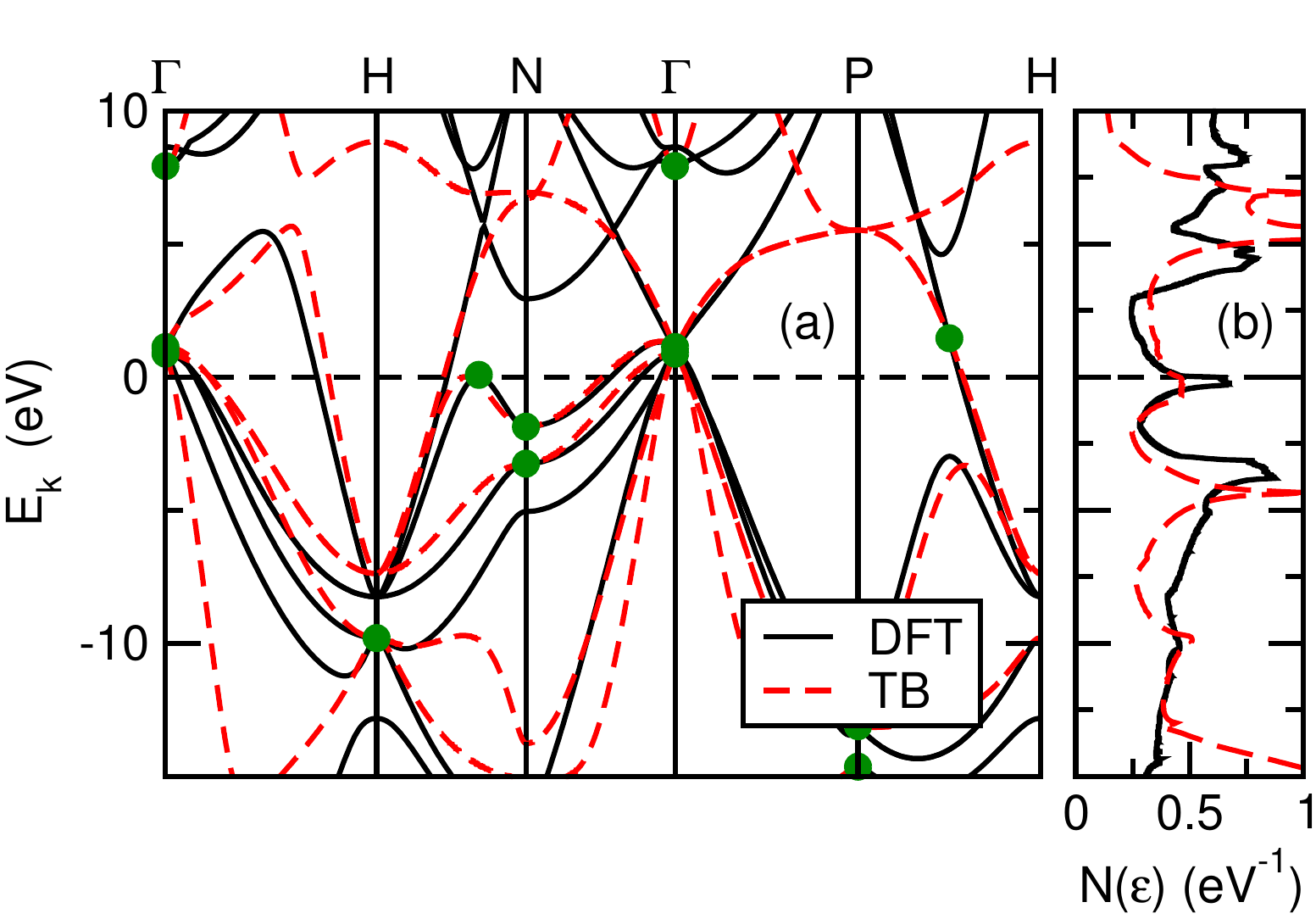}
\includegraphics[angle=0, width=8cm,clip=]{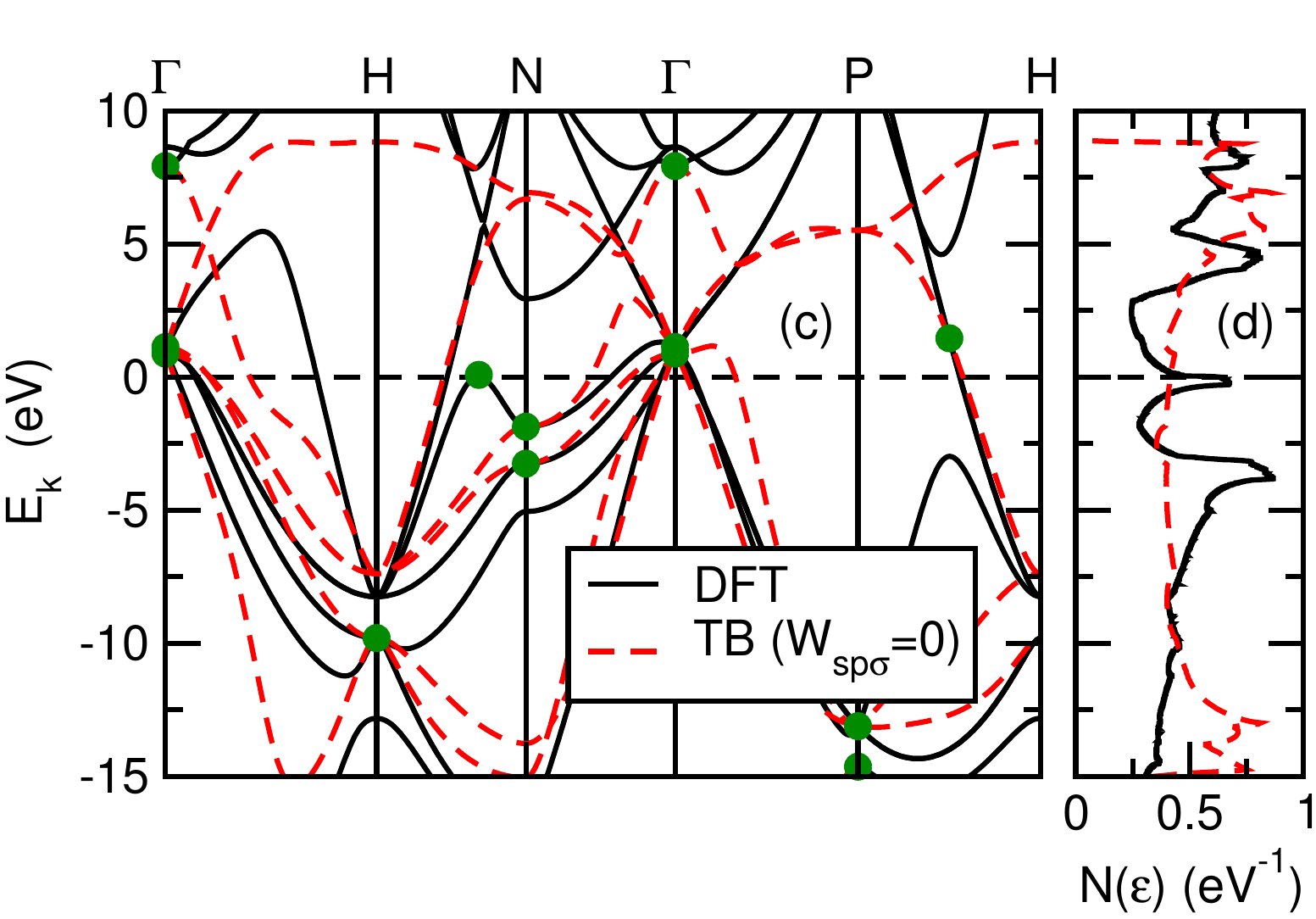}
\includegraphics[angle=0, width=7cm,clip=]{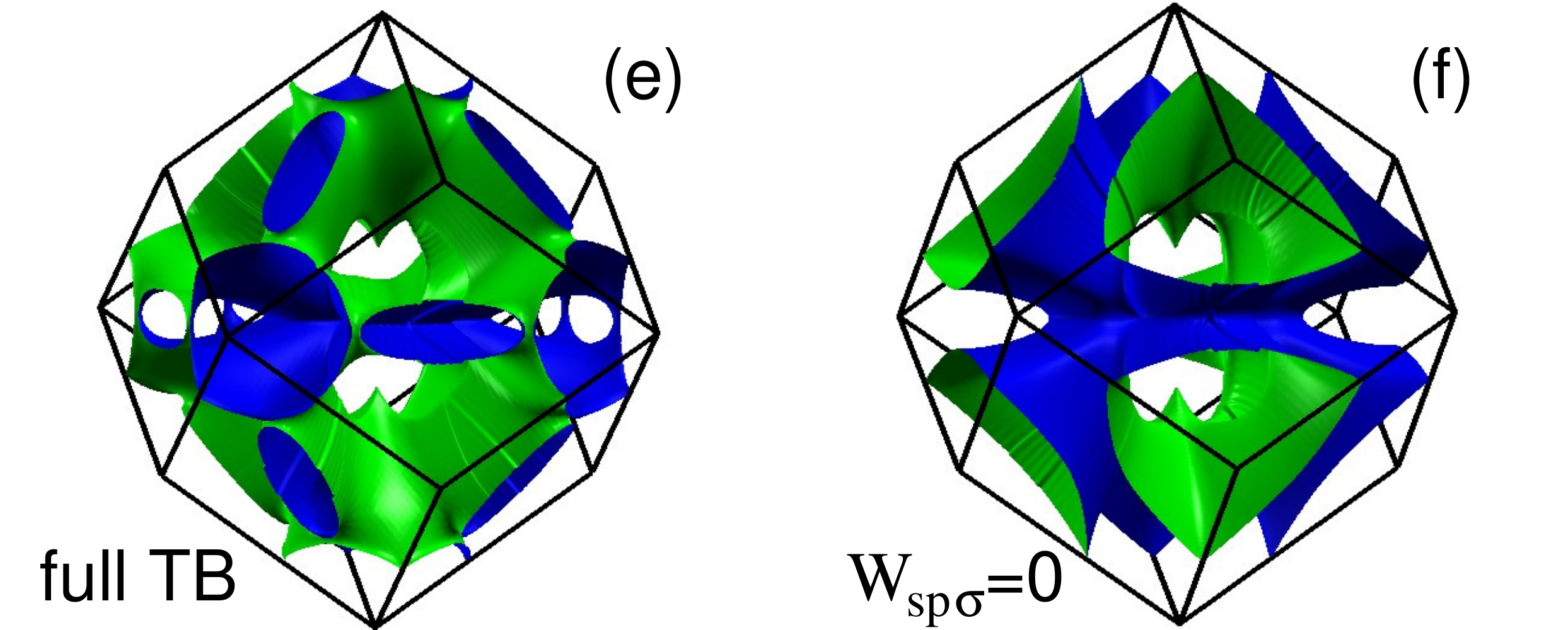}
\caption{(Color online) Panel (a): comparison between first-principle
(solid black lines) and tight-binding (dashed red lines) band
  structure. Filled green symbols represent the constraints used to determine
the tight-binding parameters. Panel (b): comparison between the corresponding
density of states. The black dashed lines mark the position of the Fermi level E$_{\text{F}}$. Panels (c)-(d): similar as panels (a)-(b) but setting $W_{sp\sigma}=0$ in the tight-binding model.
Panels (e)-(f): Fermi surfaces in the tight-binding model using the full set of TB parameters (panel e), and setting $W_{sp\sigma}=0$ (panel f).}
\label{f-bandapp}
\end{center}
\end{figure}

Motivated by the relevance of the Van Hove singularity
on the electronic properties at the Fermi level,
we choose to fix this last parameter such to reproduce the energy
of upper VHs, close to the Fermi level.
As discussed in the main text, the saddle point associated with such VHs is visible
as a local maximum in the band dispersion along the cut H-N.
The energy of such thus singularity in the our DFT calculations is
$E_{\rm VHs}^{\rm DFT}=0.085$ eV, in agreement with other DFT calculations based
on the same code.\cite{Pickett2}
We thus fix the value of the parameter $W_{sp\sigma}$ by requiring the tight-binding
band-structure to have a similar Van Hove singularity at the {\em same} energy, i.e.
$E_{\rm VHs}^{\rm TB}=0.085$ eV.
All the tight-binding parameters, estimated in this way,
are reported in Table \ref{t-TBparam},
and the comparison between the DFT and TB band structures and density
of states is shown
in Fig. \ref{f-bands}a,b, as well as, for commodity, in
Fig. \ref{f-bandapp}a,b on a different scale.
In Fig. \ref{f-bandapp}a we also mark the DFT points used as constraints to estimate
the TB parameters.
As already discussed in the main text, the overall agreement is
remarkably good.

Note that no direct information about the Fermi level was employed to
estimate the tight-binding parameters.
In this perspective, the agreement between the TB and DFT Fermi
surfaces (Fig. \ref{f-bands}c,d) is particularly significant and it points out the robustness of the present tight-binding model.

It is also worth to stress here that the direct hopping between S $3s$
and S $3p$ orbitals on nearest neighbor sulphur atoms,
governed by the parameter $W_{sp\sigma}$, is fundamental
in order to reproduce the correct low-energy properties
(and hence the Fermi properties) of the band structure.
This is indeed evident in Fig. \ref{f-bandapp}c,d
where we compare the DFT band structure and density of states
with the tight-binding model where we set $W_{sp\sigma}=0$.
As clear by construction,  the resulting TB dispersion still
reproduces the chosen DFT energy levels on the high-symmetry points,
but the overall band structure is nevertheless quite different.
More striking is the lack of the saddle point (and Van Hove
singularity) along the cut H-N, and as a consequence
the low-energy properties and the Fermi surface 
(Fig. \ref{f-bandapp}e,f)
appear radically different.
The lack of the VHs close to the Fermi level also implies of course
the lack of a peak at the Fermi level in the density of states
(Fig. \ref{f-bandapp}d).

\begin{figure}[t!]
\begin{center}
\includegraphics[width=8cm,clip=]{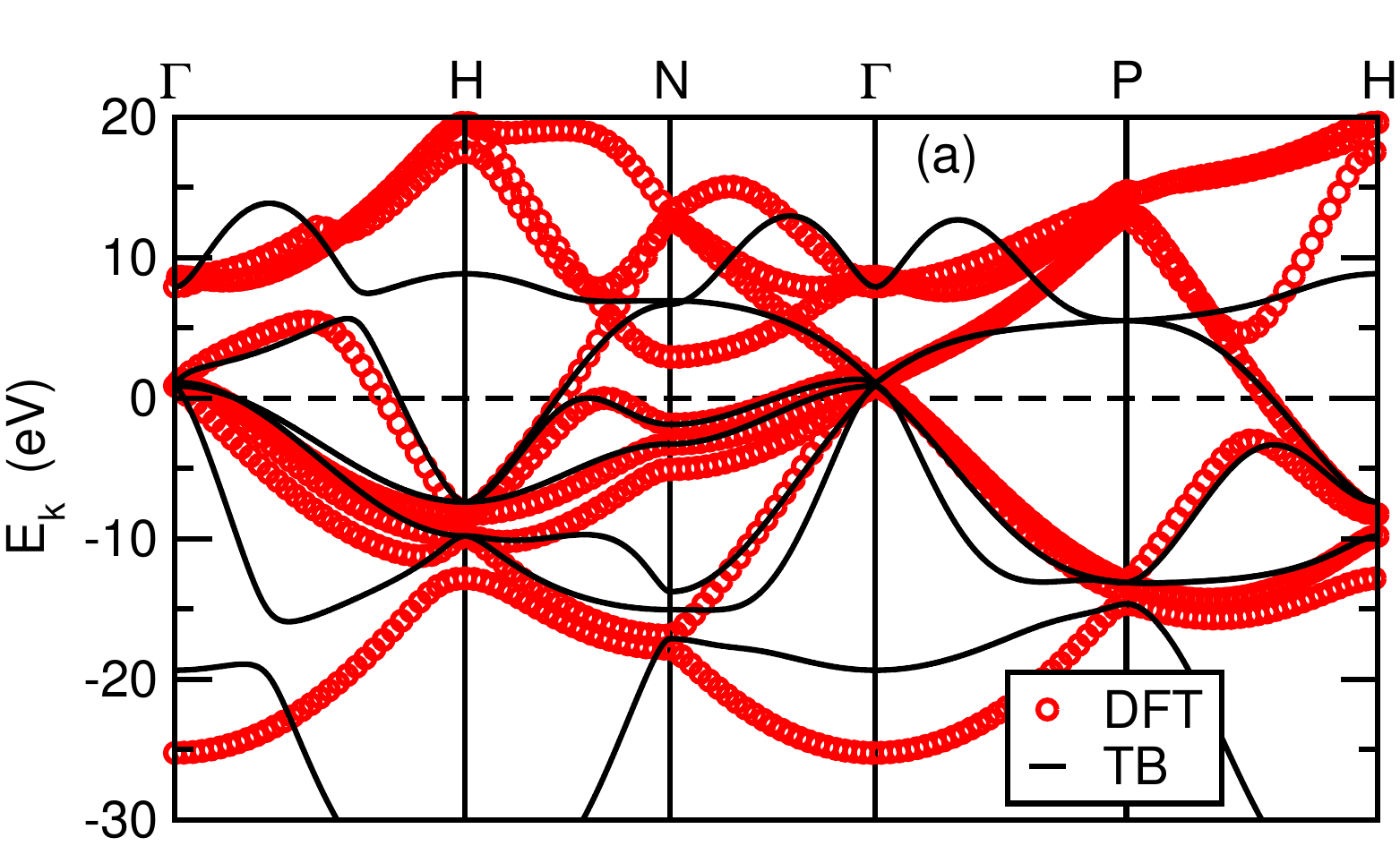}
\includegraphics[width=8cm,clip=]{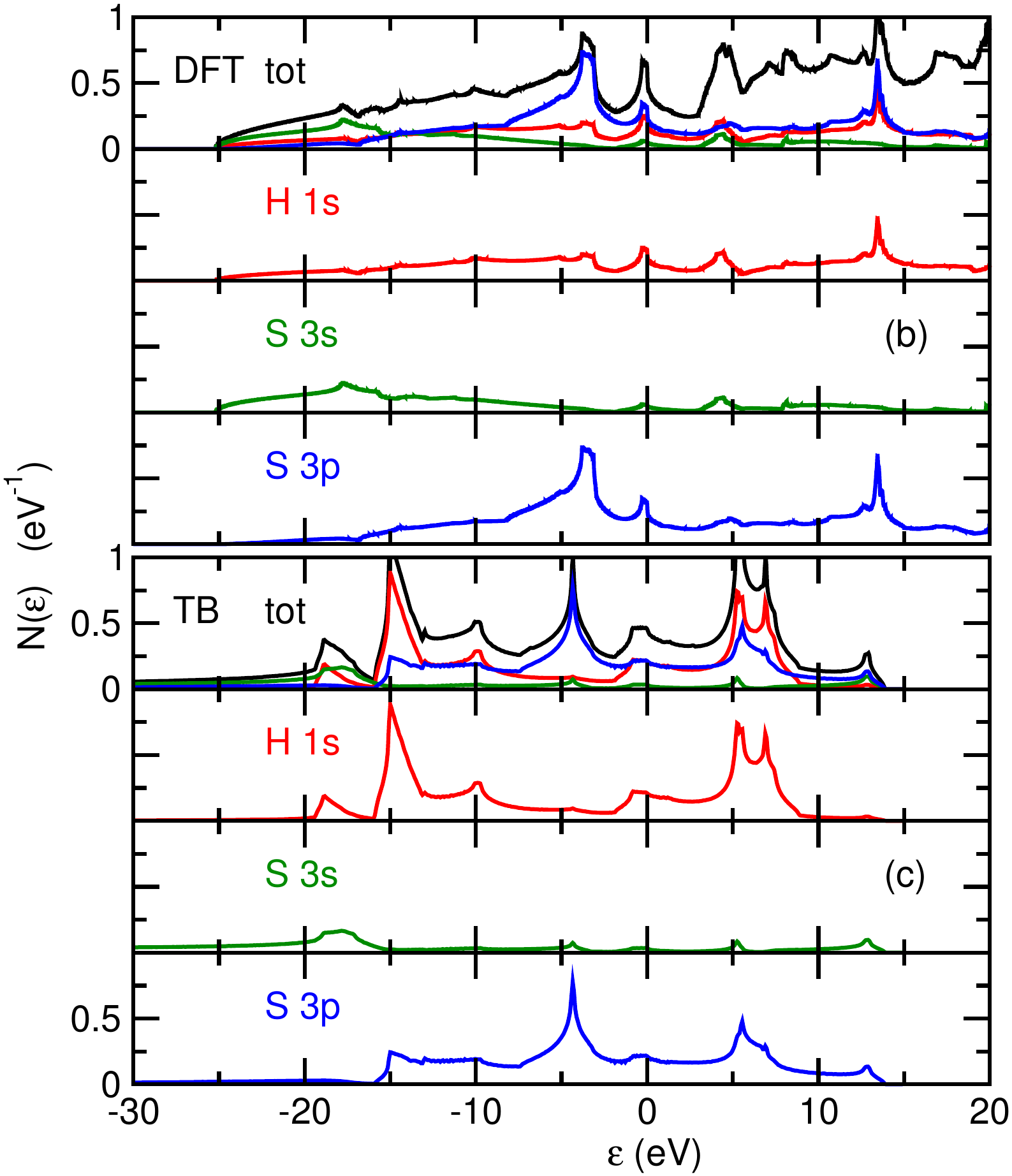}
\caption{(Color online) Panel (a): comparison between DFT and TB
band structures on a wide energy scale. Panels (b)-(c): total and partial DOS
as computed by DFT and in the tB model. In both cases, the partial DOS
is includes in the panel with the total DOS, and repeated
in the lower panels, for a better readibility.
}

\label{f-enwide}
\end{center}
\end{figure}

It is also instructive to compare the first-principle and the
tight-binding results on a wider energy scale.
The comparison for the band structure is shown in
Fig. \ref{f-enwide}a.
As discussed in the main text, considered the unavoidable
approximations, the agreement is fair in the low-energy region
$[-10:5]$ eV, whereas it appears less accurate at energies far from the Fermi level,
where the hybridization with orbitals not considered in the
tight-binding model becomes more relevant.
Further information can be gained by the analysis of the partial
density of states, i.e. density of states projected on the atomic
orbitals.
This comparison in shown in Fig. \ref{f-enwide}b,c, where we plot
the total and the projected DOS for both DFT and TB calculations.
For an easier reading, we also include, together with the total DOS,
the partial density of states which are also reported in the separate
panels.
Few points are worth here being underlined:
$i$) at low energy, the TB model captures the correct orbital component,
with a dominant equal content of H $1s$ and S $3p$,
in agreement with previous works;\cite{Pickett}
$ii$) at high-energy, the TB model estimates a total bandwidth
$\approx [-20:15]$ eV
for the set H $1s$ + S $3s$ + S $3p$ (Fig. \ref{f-enwide}c), also in good agreement
with the projected density of states obtained by first-principle
calculations (Fig. \ref{f-enwide}b).


\end{appendix}


\begin{thebibliography}{999}
\bibitem{notefermilevel}
In both first-principle calculations and in the tight-binding model,
we set at zero the energy of the Fermi level.
The on-site energies listed in Table \ref{t-TBparam}
are thus relative to the Fermi level.
\bibitem{note_TB}
It is maybe for this reason that
the combination  $S_{pp}^-$,
as well as $W_{sp\sigma}$, was neglected {\em tout court}
in the whole Brillouin zone in Ref. \onlinecite{Pickett2}.
\end{thebibliography}
\end{document}